\documentclass[12pt]{article}
\usepackage{xcite}
\usepackage{xr}
\usepackage{bm}
\usepackage[normalem]{ulem}
\usepackage{tabularx}
\usepackage{afterpage}
\usepackage{longtable}
\usepackage{array}
\newcolumntype{N}{@{}m{0pt}@{}}
\makeatletter
\newcommand*{\addFileDependency}[1]{
  \typeout{(#1)}
  \@addtofilelist{#1}
  \IfFileExists{#1}{}{\typeout{No file #1.}}
}
\makeatother
\newcommand*{\myexternaldocument}[1]{%
    \externaldocument{#1}%
    \addFileDependency{#1.tex}%
    \addFileDependency{#1.aux}%
}
\myexternaldocument{web_file}
\pdfoutput=1
\usepackage{graphicx}
\usepackage{amsmath,color,epsfig,amsfonts,fullpage, natbib, rotating}
\usepackage[font={small,it}]{caption}
\usepackage{subcaption}
\usepackage{appendix}
\usepackage[colorlinks=true, pdfstartview=FitV, linkcolor=blue,
citecolor=blue, urlcolor=blue]{hyperref}
\usepackage{verbatim} 
\usepackage{ctable} 
\usepackage{amsthm}

\usepackage{amssymb}

\newcommand{\beq}{\begin{equation*}}
\newcommand{\eeq}{\end{equation*}}
\newcommand{\nch}{\left\{\begin{array}{ll}}\newcommand{\ech}{\end{array}\right.}

\def\Mg{\mathcal{M}_\gamma}
\def\Mn{\mathcal{M}_N}

\def\Mone{\mathcal{M}_1}
\def\bb{\bm{\beta}}
\def\hbb{\hat{\bb}}
\def\hbbls{\hat{\bb}_{LS}}
\def\hbbbma{\hat{\bb}_{BMA}}
\def\bbg{\bb_\gamma}
\def\bt{\bm{\theta}}
\def\btg{\bt_\gamma}

\def\gg{g_\gamma}
\newcommand{\bols}{\hat{\beta}_{LS}}
\def\bone{\mathbf{1}}

\def\bY{\bm{Y}}
\def\bbarY{\bar{\bY}}
\def\bX{\bm{X}}
\def\bXg{\bX_\gamma}
\def\bg{\bm{g}}
\def\bTh{\bm{\Theta}}

\makeatletter
\def\blfootnote{\xdef\@thefnmark{}\@footnotetext}
\makeatother

\begin{document}

\begin{center}
\section*{
Empirical Bayes Model Averaging\\
with
Influential Observations:\\ Tuning Zellner's $g$ Prior for Predictive Robustness}
\vspace{-.25cm}

\textbf{Christopher M.~Hans}$^{a,*}$,
\textbf{Mario Peruggia}$^a$ and
\textbf{Junyan Wang}$^a$

$^a$\emph{Department of Statistics, 
The Ohio State University,
Columbus, OH 43210, USA}

* Corresponding author: (hans@stat.osu.edu)


\end{center}
\vspace*{-7mm}
\begin{abstract}
  The behavior of Bayesian model averaging (BMA) for the normal linear
  regression model in the presence of influential observations that
  contribute to model misfit is investigated.  Remedies to attenuate
  the potential negative impacts of such observations on inference and
  prediction are proposed. The methodology is motivated by the view
  that well-behaved residuals and good predictive performance often go
  hand-in-hand. Focus is placed on regression models that use
  variants on Zellner’s $g$ prior. Studying the impact of various
  forms of model misfit on BMA predictions in simple situations points
  to prescriptive guidelines for ``tuning'' Zellner’s $g$ prior to
  obtain optimal predictions.  The tuning of the prior distribution is
  obtained by considering theoretical properties that should be
  enjoyed by the optimal fits of the various models in the BMA
  ensemble. The methodology can be thought of as an ``empirical
  Bayes'' approach to modeling, as the data help to inform the
  specification of the prior in an attempt to attenuate the negative
  impact of influential cases.

\end{abstract}

\vspace{-2mm}
\noindent
KEY WORDS: Bayesian regression; BMA; Outliers; Regression diagnostics; Residuals; Shrinkage estimation.

\vspace{-6mm}
\section{Introduction}\label{sec:intro}

While there are many strategies for Bayesian regression modeling when
there is uncertainty about model composition, Bayesian model averaging
has been predominant in both literature and practice over the last two
decades.  This is particularly true when prediction is the primary
objective.  In this setting, a prior distribution is first placed over
the collection of all models under consideration.
For each model $\Mg$, a prior distribution is then
placed on the model-specific parameters, leading to a joint posterior
distribution over both models and parameters.  Predictions of a new
outcome $\tilde{\bY}$ are typically made by computing a weighted average
of the predictions under all possible models,
\[
\mbox{$\mbox{Pred}(\tilde{\bY}) = \sum_{\gamma \in \Gamma} 
\mbox{Pred}(\tilde{\bY} \mid \Mg) w(\Mg),$} 
\]
where $\Gamma$ is a set that indexes all models under consideration,
$\Mg$ is a particular model and $w(\Mg)$ is the weight assigned to
model $\Mg$ in the averaging.  Most commonly, $\mbox{Pred}(\tilde{\bY})$
is taken to be $E(\tilde{\bY} \mid \bY)$, which minimizes expected
posterior predictive $L_2$ loss, in which case the model-specific
predictions are $E(\tilde{\bY} \mid \bY, \Mg)$ and the
model-specific weights are the posterior probabilities $\pi(\Mg \mid
\bY)$ \citep{bern:94}.  Reviews of BMA are provided in
\cite{hoet:99} and \cite{clyd:04}.
The term ``BMA'' is quite general and can be applied to many different 
data-analytic settings. In this work, ``BMA'' refers to Bayesian model averaging 
for linear regression, where an analyst has available $p$ potential covariates
and would like to average over the $2^p$ possible regression models
that can be constructed using subsets of the covariates. Seminal work on
regression model composition uncertainty, which underlies BMA, 
can be found in \cite{mitc:88}, \cite{geor:93}, \cite{smit:96}, \cite{geor:97},
\cite{raft:97} and \cite{chip:01}.

While methods for diagnosing and remedying model misfit are well-established 
and routinely taught in regression modeling courses 
\citep[see, e.g.,][]{nete:96, cook:82}, the literature on model misfit for Bayesian 
analyses of classical linear regression models is somewhat less well-developed. 
Addressing potential or known model misfit is sometimes accomplished by modifying 
the likelihood, writing down a model that is flexible enough to describe the observed 
data. One example is the use of heavy-tailed error distributions to accommodate outliers 
\citep{west:84}. Model-based approaches to outlier detection and residual analysis have 
been considered by \citet{chal:88} and \citet{hoet:96}. 

The corpus of literature on Bayesian regression modeling and, in particular, Bayesian 
averaging of many regression models has expanded greatly since the development of
early work addressing model misfit in Bayesian regression. While existing work provides 
useful guidelines for thinking about model misfit for specific Bayesian models, it is not clear 
that it provides prescriptive guidelines that can be applied in many of the currently-used 
Bayesian regression modeling settings. 

For example, priors related to Zellner’s $g$ prior \citep{zell:86} have become one of the 
``standard'' setups for Bayesian regression modeling. The popularity of such priors is due, 
in part, to their computational simplicity. Use of these priors requires choosing an approach 
for handling hyperparameters, which we refer to generically as $g$. Current popular approaches for 
handling $g$ include empirical Bayes (EB) methods that focus on specifying $g$ by maximizing 
the marginal likelihood of the data with respect to either a single model or a mixture of models. 
Fully Bayes approaches assign a prior distribution to $g$ and then integrate it out of the model.
While these approaches are sound when the model fits well, it is less clear that maximizing 
the marginal likelihood or integrating $g$ out of the model will be optimal in the presence of model 
misfit due to influential outliers, as the discrepancy between prior and likelihood may result in 
sub-optimal out-of-sample inference (say, the prediction of new cases). 

In this work, we focus on methods of model specification in the presence of influential outliers that are 
strongly connected to the concept of residual analysis. We take the view that well-behaved residuals and 
good predictive performance usually go hand-in-hand. Any particular choice for handling $g$ corresponds 
to a specific Bayesian model and hence specific fitted values and residuals.
By tuning the value of $g$ in Zellner's $g$ prior over a continuous interval, we obtain 
continuously-varying residuals. Broadly speaking, in the presence of model misfit we seek to choose a 
value of $g$ that yields well-behaved residuals and hence attenuates the impact of the model misfit and, 
ultimately, achieves better out-of-sample predictive performance. In practice, we seek to find prescriptive 
guidelines for accomplishing this task by focusing on minimizing prediction error. Our investigation concerns 
not only an individual regression model but also an entire ensemble of regression models that may be combined 
together via BMA. In this case, model misfit due to influential outliers can exist at both the local (individual model) 
level and the global (model-averaged) level. By tuning the priors to attenuate the impact of model misfit, we 
eventually achieve improved out-of-sample predictive performance.

In Section~\ref{sec:regression_model} we introduce our regression model setting and review Zellner's
$g$ prior. Section~\ref{sec:intuition} makes connections between the choice of $g$ and optimal prediction
in the presence of influential outliers, and provides prescriptive guidelines for tuning Zellner's $g$ prior.
Section~\ref{sec:methods} describes the approach we have developed for applying these prescriptive guidelines in 
practice, and Section~\ref{sec:sims} investigates the performance of this approach in simulation studies.
Section~\ref{sec:crime} applies our approach in an analysis of a data set and compares its predictive
performance to other common methods.
A summary of the results and a discussion of open questions and related work are provided in 
Section~\ref{sec:discussion}.

\section{Regression model setting}\label{sec:regression_model}

We consider situations where an analyst observes an $n \times 1$ vector
of response values, $\bY$, and an $n \times p$ matrix $\bX$ which contains 
$p$ covariate values for each of the $n$ cases. As is common in the literature,
unless specified otherwise, we assume throughout that the columns of $\bX$ 
have been mean centered \citep[see, e.g.,][for justifications and discussion 
of this modeling choice]{lian:08, baya:12, li:18}. There are $2^p$ possible subsets 
of the $p$ covariates that could be used to construct a mean function for a linear 
regression model, and interest lies in computing model-averaged predictions as 
described in Section~\ref{sec:intro} over the space of all possible regression models. 
In this setup, a model $\Mg$ corresponds to a subset of $k$ predictors.
We use the binary $p$-vector $\gamma$ to index all possible models: $\gamma_j = 1$
when $X_j$ is included in a model and $\gamma_j = 0$ otherwise. The notation
$\Mg$ is used to denote a model containing predictors indicated by the vector $\gamma$.
Other quantities subscripted by $\gamma$ denote model-specific terms, e.g.~$\bXg$ 
is the $n\times k$ matrix which contains the columns of $\bX$ that are included in
model $\Mg$.

Our work focuses on data analysis settings where either (i) the total number
of predictors $p$ is not too large or (ii) where $p$ may be large, but we constrain the maximum 
number of predictors in a model to be $k^* \ll p$ \citep[e.g., as in][]{hans:07}.
The simulations and examples presented in Sections~\ref{sec:sims} and \ref{sec:crime}
consider $p$ as large as ten. We focus on small-$p$ settings in part because
they are relevant for a wide swath of applied data analysis, where an investigator is interested in 
understanding the relationship between a small set of regressors and a response variable, and
in part due to the computational challenges associated with averaging over very large model
spaces that are faced by all BMA methods.

BMA requires hierarchical specification of a prior probability, $\pi(\Mg)$, for each
model under consideration, and then, for each model, a likelihood and a
prior for the model-specific regression parameters. For normal linear regression
modeling, the likelihood for a specific model $\Mg$ is derived from the distribution of 
$\bY$ given the model parameters, $(\alpha, \bbg, \sigma^2)$,
\[
	p(\bY \mid \alpha, \bbg, \sigma^2, \Mg) = 
		\mbox{N}(\bY \mid \alpha \mathbf{1}_n + \bXg \bbg, 
			\sigma^2 I_n),
\]
where $\bbg$ are the regression coefficients
corresponding to the predictors $\bXg$ in model $\Mg$, $\alpha$
is an intercept, $\sigma^2$ is the error variance, $\bone_n$ is 
an $n \times 1$ vector of ones, and $I_n$ is the $n \times n$ identity
matrix. The model-specific priors on the parameters are known
to play a key role in inference and prediction, and there are many possibilities
for their specification. A common approach, which we adopt in this work,
is to assume $\pi(\alpha, \bbg, \sigma^2 \mid \Mg) = 
\pi(\bbg \mid \sigma^2, \Mg)\pi(\alpha, \sigma^2
\mid \Mg)$ and to use the improper, ``objective'' prior $\pi(\alpha, \sigma^2
\mid \Mg) \propto \sigma^{-2}$ \citep{jeff:61, berg:98}.

Various approaches to specifying the prior probabilities $\pi(\Mg)$ have been studied in the 
literature. While the uniform prior over models, $\pi(\Mg) = 2^{-p}$, was discussed in the early 
Bayesian variable selection literature \citep{geor:93}, it does not control the false positive rate in 
the context of variable selection when making multiple comparisons \citep{scot:10}.
An alternate prior formulation that provides
better multiplicity control is the $\mbox{Beta-Binomial}(1,1)$ prior, $\pi(\Mg) =
(p+1)^{-1} {p \choose |\Mg|}^{-1}$, where $|\Mg|$ is the number of variables in model $\Mg$
\citep{ley:09, scot:10}. This is a special case of the Beta-Binomial prior considered by \citet{kohn:01}
and can be interpreted as first specifying a uniform prior over model size and then, conditionally
on model size, specifying a uniform prior over models. More recent advances in model space
prior specification include the loss-based prior of \citet{vill:20}. Unless specified otherwise, we
use the $\mbox{Beta-Binomial}(1,1)$ prior in this work.

The research literature on priors for the model-specific regression coefficients, 
$\pi(\bbg \mid \sigma^2, \Mg)$, is extensive. Historically, one of the most popular classes 
of priors for the regression coefficients is based on Zellner's $g$ prior,
\begin{equation}
	\bbg \mid \sigma^2, \Mg \sim \mbox{N}(0, \gg \sigma^2 (\bXg^T \bXg)^{-1}),
		\label{eq:gprior}
\end{equation}
where $\gg > 0$ is a hyperparameter \citep{zell:86}. This prior has received much attention
in the literature and in practice due in part to the fact that, for specific treatments of 
$\gg$, Zellner's $g$-prior allows for computationally efficient model averaging when $p$ is not 
too large. In the context of BMA, when minimizing expected posterior predictive $L_2$ loss, 
model averaged predictions require calculation of $E(\bbg \mid \bY, \Mg)$ and 
$\pi(\Mg \mid \bY) \propto m(\bY \mid \Mg) \pi(\Mg)$ for each model, where $m(\bY \mid \Mg)$ 
is the marginal likelihood for model $\Mg$. Closed form expressions for these quantities exist 
when $\gg$ is treated as a constant in the model. The marginal likelihood for model $\Mg$ is
\begin{eqnarray*}
	m(\bY \mid \Mg) &=& \int p(\bY \mid \alpha, \bbg, \sigma^2, \Mg)\pi(\alpha, \bbg, \sigma^2 \mid \Mg) 
		\, d\alpha \, d\bbg \, d\sigma^2 \\ 
		&=& \frac{\Gamma((n-1)/2)}{\pi^{(n-1)/2} n^{1/2} |\!| \bY - \bbarY |\!|^{n-1}}   
			\frac{(1 + \gg)^{(n-k-1)/2}}{(1 + \gg(1 - R_\gamma^2))^{(n-1)/2}},
\end{eqnarray*}
where $\bbarY$ is the $n\times 1$ vector where all elements are equal to the sample average value of 
the $Y_i$'s, $|\!| \cdot |\!|$ is the $L_2$ norm and $R^2_\gamma$ is the coefficient of determination 
for model $\Mg$. The posterior mean of the regression coefficients under model $\Mg$ is
\begin{equation}
	E(\bbg \mid \bY,  \Mg) = \frac{\gg}{1+\gg} \hbbls,
	\label{eq:bshrink}
\end{equation}
where $\hbbls$ is the ordinary least squares estimate of $\bbg$ under model $\Mg$, and
$\rho_\gamma = \gg/(1+\gg) \in (0,1)$ is sometimes called the \emph{shrinkage factor}. Under this model, the prior 
combines with the data to shrink the least squares estimate toward zero by a factor of $\rho_\gamma$. 
Computation of all of these quantities is trivial once the usual calculations for least-squares estimation 
have been performed.

Specification (or modeling) of $\gg$ is important, as it impacts the
marginal likelihood, which in turn impacts model averaging. 
There is a large literature discussing methods for handling this parameter.
Approaches that specify fixed values of the $\gg$ include the unit information prior
\citep{kass:95}, the risk inflation criterion \citep{fost:94},
the local empirical 
Bayes prior \citep{hansen:01} and the global empirical Bayes prior
\citep{george2000calibration, clyd:00}.  Some of these approaches specify a single
$g$ that is common to all models, while others specify different $\gg$'s
for different models.  \cite{som:14} use different $\gg$'s for different groups
of predictor variables. 
\cite{lian:08} and \cite{baya:12} provide motivation
for assigning a prior distribution to $g$ and suggest priors that
enjoy appealing theoretical 
properties while maintaining computational tractability.  Other priors
for $g$ have 
been proposed by \cite{zell:80}, \cite{west:03}, \cite{cui:08},
\cite{maru:10}, and \cite{maru:11}.

The method we describe in Section~\ref{sec:methods} can be viewed as an empirical
Bayes approach to specifying hyperparameter values, and so existing empirical
Bayes methods are of special interest to us. Global empirical Bayes methods
use the data to estimate a hyperparameter~$g$ that is common to all models. The 
parameter is typically estimated by maximizing
the marginal likelihood of the observed data given $g$, $m(\bY \mid g) = 
\sum_{\gamma \in \Gamma} m(\bY \mid \Mg, g)\pi(\Mg)$, where the parameters
$\alpha$, $\bbg$ and $\sigma^2$ have been integrated out of each model to obtain
the model-specific marginal likelihood $m(\bY \mid \Mg, g)$. \citet{clyd:01} describes
an EM algorithm that can be used to find the value of $g$ that maximizes the model-averaged
marginal likelihood. The local empirical Bayes approach allows
for different $\gg$ for each model and uses the data to estimate them by maximizing
the model-specific marginal likelihood $m(\bY \mid \Mg, \gg)$. While empirical Bayes
approaches use the data to estimate specific values $\gg$, fully-Bayesian approaches
assign prior distributions to the $\gg$ and use the data to integrate over uncertainty
about them \emph{a posteriori}. Of the fully-Bayes methods, the hyper-$g/n$ prior
of \citet{lian:08} is noteworthy as it is amenable to efficient computation and has
desirable theoretical properties. This approach assigns the prior $\pi(\gg \mid \Mg) =
(a-2)(1 + \gg/n)^{-1/2}/(2n)$, with $a = 3$ recommended as a default specification.
We compare our method to these three approaches in Section~\ref{sec:sims}.

The prior mean vector $E(\bbg \mid \sigma^2, \Mg)$ is routinely taken to be the
zero vector in (\ref{eq:gprior}). When lacking prior information about the location
of the vector of regression coefficients, centering the prior at zero shrinks the posterior
distribution toward zero---corresponding to no linear association between $\bY$ and the
collection of predictors $\bXg$---as can be seen in~(\ref{eq:bshrink}). Zellner's original
formulation of the $g$ prior \citep{zell:86} allowed for shrinkage toward non-zero mean 
vectors. In this case, we write the prior as
\begin{equation}
	\bbg \mid \sigma^2, \Mg \sim \mbox{N}(\btg, \gg \sigma^2 (\bXg^T \bXg)^{-1}),
		\label{eq:nzgprior}
\end{equation}
where $\btg$ is the prior mean vector for $\bbg$. Under this prior, the posterior mean of
$\bbg$ under model $\Mg$ is
\[
	E(\bbg \mid \bY, \Mg) =  \frac{1}{1+\gg}\btg + \frac{\gg}{1+\gg} \hbbls =
		(1-\rho_\gamma)\btg + \rho_\gamma \hbbls,
\]
where $\rho_\gamma = \gg/(1+\gg)$, called the shrinkage factor above, can now be interpreted 
as determining the posterior mean as a convex combination of the prior mean of $\bbg$ and the 
data-based estimate of $\bbg$. The marginal likelihood under prior (\ref{eq:nzgprior})
for a model $\Mg$ with $k$ predictors is
\begin{eqnarray}
	m(\bY \mid \Mg) &=& \frac{\Gamma((n-1)/2)}{\pi^{(n-1)/2} n^{1/2}} |\! | \bY - \bbarY |\! |^{-(n-1)}
			(1 + \gg)^{(n-k-1)/2} \nonumber \\
	&& \hspace{1in} \times \left(1 + \gg (1 - R_\gamma^2) + \frac{|\!| \bY - \bXg \btg |\!|^2 - |\!| \bY |\!|^2}
				{|\!| \bY - \bbarY |\!|^2}\right)^{-\frac{n-1}{2}}.
			\label{eq:nzmarglik}
\end{eqnarray}
Dividing (\ref{eq:nzmarglik}) by the marginal likelihood for the ``null model'', $\Mn$---the model with no 
predictors that assumes a common mean for all $Y_i$---yields the Bayes factor for comparing model $\Mg$ 
to model $\Mn$:
\begin{equation}
	BF(\Mg : \Mn) = (1+g_\gamma)^{(n-k-1)/2} 
		\left(1+g_\gamma(1-R_\gamma^2) + \frac{||\bm{Y}-\bm{X_\gamma\theta_{\gamma}}||^2-||\bm{Y}||^2}
		{||\bm{Y}-\bm{\bar{Y}}||^2}\right)^{-\frac{n-1}{2}}. \label{eq:nzbf}
\end{equation}
We revisit this version of Zellner's $g$ prior in Section~\ref{sec:nonzerog}.

\section{Outliers and Model-Averaged Prediction Accuracy}\label{sec:intuition}

Model-averaged predictions for linear regression depend on the posterior distribution over 
models, $\pi(\Mg \mid \bY)$, and the within-model estimates of the regression coefficients, 
$E(\bbg \mid \bY, \Mg)$, both of which depend on $\gg$. When $p$ predictors are available
and we are averaging over the space $\Gamma$ of all possible $2^p$ models, the 
model-averaged predictions of $n$ new cases $\tilde{\bY}$ at the same matrix of regressors
$\bX$ used to fit the model can be written as
\begin{eqnarray*}
	\mbox{Pred}(\tilde{\bY}) &=& \sum_{\gamma \in \Gamma} \mbox{Pred}(\tilde{\bY} \mid \Mg)
		\pi(\Mg \mid y) \\
	&=& \sum_{\gamma \in \Gamma} \left(\bbarY + \bXg E(\bbg \mid \bY, \Mg)\right) 
		\pi(\Mg \mid \bY) \\
	&=& \bbarY + \bX E(\bb \mid \bY),
\end{eqnarray*}
where, for $j = 1, \ldots, p$, the $j$th element of the vector $E(\bb \mid \bY)$,
\[
	E(\beta_j \mid \bY) = \sum_{\gamma \in \Gamma \; : \; X_j \in \Mg} 
		E(\beta_{\gamma, X_j} \mid \bY, \Mg) \pi(\Mg \mid \bY),
\]
is the model-averaged estimate of $\beta_j$, and the sum is taken over all
models that include $X_j$ as a regressor. The notation $\beta_{\gamma, X_j}$
is used to denote the coefficient in model $\Mg$ that corresponds to regressor $X_j$.
We denote the model-averaged estimate of
the regression coefficients as $\hbbbma \equiv E(\bb \mid \bY)$.

In this section, we present several examples to illustrate the impact of influential outliers on BMA.
In particular we focus on the relationship between outlying value(s), the choice of $\gg$ in the 
prior, and predictive accuracy. The examples shed light on the behavior of
BMA in the presence of influential outliers and suggest general strategies for improving 
model-averaged prediction.
Example~1 illustrates the sensitivity of the posterior distribution over models to outliers
in an example with three potential predictors. Example~2 focuses on the impact of
outliers on model-averaged prediction in a stylized example. This example makes
connections between the behavior of residuals and optimal prediction. 
Sections~\ref{sec:meanshift} and \ref{sec:varianceinflation} generalize these
results to mean-shift and variance-inflation contamination settings with an
arbitrary number of predictors, $p$. The examples and results in this section
provide the intuition behind the prescriptive guidelines for selecting $\gg$ that we pursue
in Section~\ref{sec:methods}.

\subsection{Example~1: Impact of Outliers on the Posterior}\label{sec:ex1}

We first present a simple example that illustrates the impact of a single contaminated
case on the posterior distribution over models.  Let $X_{i1} = i - 1 - (n-1)/2$,
for $i= 1, \ldots, n = 11$, so that the $X_{i1}$ take the values $-5, -4, \ldots, 4, 5$ and
are mean-centered with $\sum_{i=1}^n X_{i1} = 0$. Suppose that the true regression line is 
given by $E (Y_i  \mid  X_i) = \beta X_{i1}$ and that the data are generated according to
\[
Y_i = \beta X_{i1} + I(i=j)K  + \epsilon_i, \,\, i = 1, \ldots, n,
\]
with $\epsilon_i \stackrel{iid}{\sim} {\rm N}(0,1)$ and 
$\beta =0.5$, where $I(i = j)$ is a 0/1 indicator function that is nonzero 
only for case $j$. In this setup, $K\neq 0$ means the response for case $j$ is a
contaminated value and constitutes an outlier if $|K|$ is large.
Case $j$ will also have high leverage for small and large values of
$j$.  We simulated $(X_{i1}, Y_i)$ pairs with $n = 11$, $\beta =
0.5$ and $\sigma^2 = 1$.  We also simulated (and then mean centered) two other 
potential predictor variables: $X_{i2} = X_{i1} + Z_i$, with 
$X_i \stackrel{iid}{\sim} {\rm N}(0,1)$ so that $X_1$ and $X_2$ are highly 
correlated, and $X_{i3} \stackrel{iid}{\sim} {\rm N}(0,1)$ with $X_3$ 
independent of $Y$, $X_1$ and $X_2$. Considering all three potential 
predictors, the model space contains $2^3 = 8$ possible models (including 
the null model, containing only an intercept), over which we place a prior distribution 
that is uniform over model size and, conditional on model size, uniform over models.
For each of the non-null models, the prior distribution on the
regression coefficients is taken to be Zellner's $g$-prior centered
at zero with $g = n$, corresponding to a unit information prior 
\citep{kass:95, fern:01}. The prior is completed with the standard
non-informative prior $\pi(\alpha, \sigma^2) \propto \sigma^{-2}$. 

\begin{figure}[t]
\centering
	\includegraphics[width=1.0\linewidth]{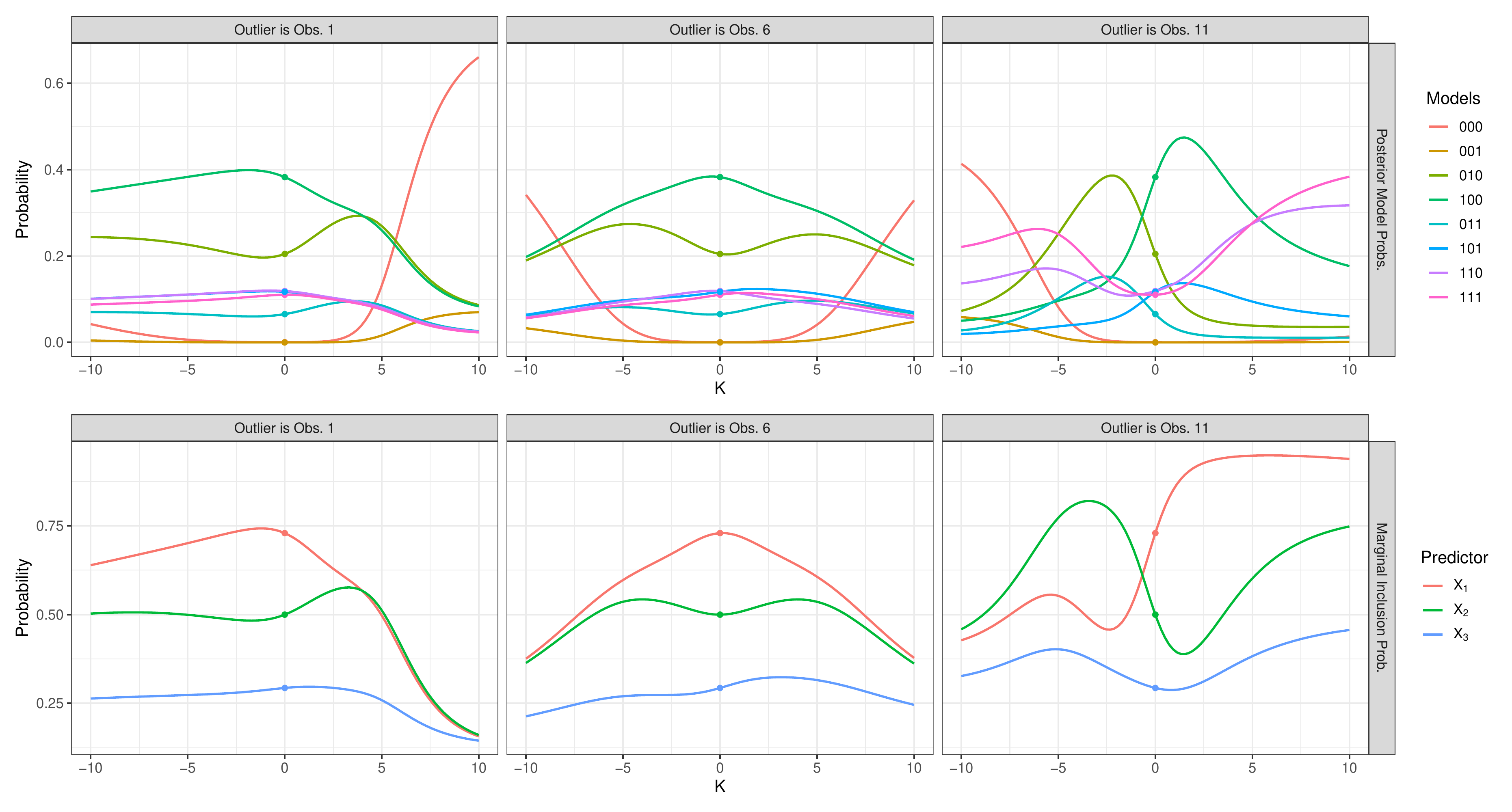}
         \caption{Posterior model probabilities (top row) and marginal inclusion
           probabilities (bottom row) for Example~1 in Section~\ref{sec:ex1}.
           Models in the top row are indexed by $\gamma$, e.g.~``010'' indicates the
           model with only $X_2$ as a predictor.}
           \label{fig:modprobs}
\end{figure}

For a given contaminated case location $j$, we are interested in the
impact of $K$ on the  posterior probabilities of the eight models. The top
row of Figure~\ref{fig:modprobs} displays this impact as $K$ ranges from $-10$ to 
$10$ for $j = 1$ (so that the contaminated case corresponds to the smallest $X_1$
value), $j = 6$  
(medium $X_1$ value) and $j = 11$ (largest $X_1$ value). The binary vector $\gamma$ distinguishes
the eight models: $(0,0,0)$ corresponds to the model
that includes none of the three predictors, $(1,0,0)$ contains only $X_1$,
$(0,1,1)$ contains $X_2$ and $X_3$, etc. Clearly, the posterior distribution is quite
sensitive to the contaminated case.  In cases where the contaminated
case attenuates the 
relationship between $X_1$ and $Y$ ($j\approx 1$, $K \approx 10$;
$j\approx 11$, $K \approx -10$),  
the null model becomes more heavily weighted than when there is no
contaminated case. 
In other cases (e.g., $j=11$, $K\approx 10$), the posterior shifts
toward models  
that include too many predictors. When $K=0$ there is no outlier, and so the
posterior model probabilities will be the same in each plot ($j = 1, 6, 11$) as
indicated by the plotted points.

The marginal inclusion probabilities for each predictor $j$,
\[
	\pi(\gamma_j = 1 \mid \bY) = \sum_{\gamma \in \Gamma \; : \; \gamma_j=1}
		\pi(\Mg \mid \bY),
\]
play an important role in assessing model uncertainty from a Bayesian perspective. 
\citet{barb:04} provide conditions under which
the \emph{median probability model} (MPM)---the model that includes all
predictors with $\pi(\gamma_j = 1 \mid \bY) \geq 0.5$---is Bayes-optimal for
prediction when a single model is to be selected. \citet{carv:08} provide
decision-theoretic support for reporting the MPM under a model selection
framework. Marginal inclusion probabilities also play a key role in many
computational approaches for BMA \citep[e.g.,][]{clyd:11}.

The bottom row of Figure~\ref{fig:modprobs} displays how outlier location,
magnitude and direction impact the marginal inclusion probabilities for
each of the three predictors in this example. The probabilities are clearly
sensitive to the nature of the outlier. When $K=0$ (no outlier), the MPM just
barely selects the generative model ($X_1$ only); the marginal inclusion probability
for $X_2$ is $0.4996$. Outlier contamination suppresses marginal inclusion probabilities
in some cases (outlier location $j = 1$, $K$ large; $j = 6$, $|K|$ large) while amplifying
them in others ($j=11$, $K$ large).
Model-averaged inference and prediction depend on the posterior weights
for the individual models, which we learn from this example can be impacted 
greatly by the presence of an influential outlier. This motivates the need to study the 
impact of outliers on BMA predictive accuracy, which we explore in the next example.

\subsection{Example 2: Impact on Prediction}\label{sec:ex2}
\begin{figure}[b!]
\centering
         \includegraphics[scale = 0.55]{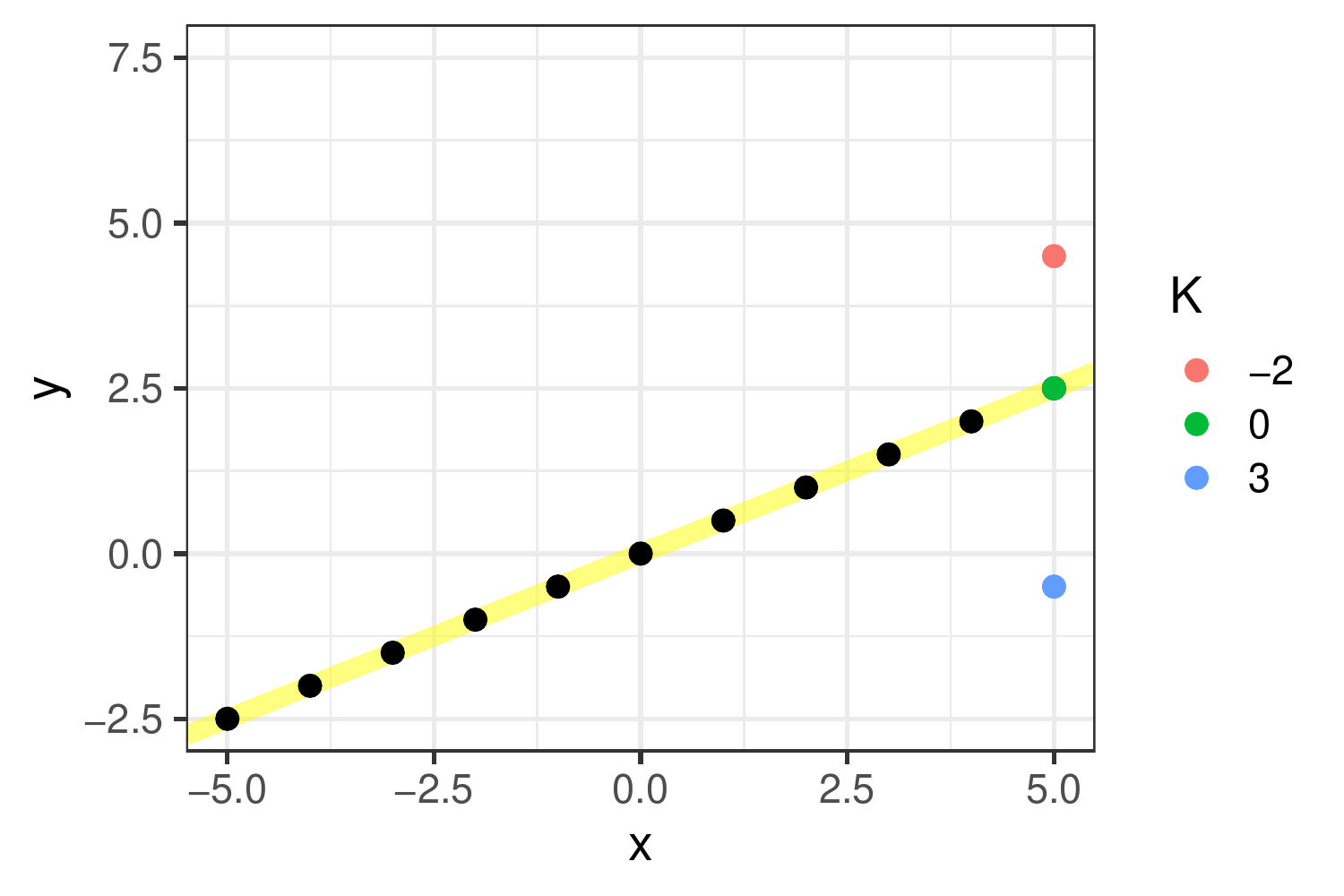}\\
         \caption{Simple linear regression data with
           contaminants for Example~\ref{sec:ex2}}\label{fig:slr_contaminants}
\end{figure}

This example considers a simplified version of Example~1 in order to illustrate
clearly how the choice of $g$ in the $g$ prior impacts model averaged prediction
accuracy in the presence of an influential outlier. In this example we consider a single
potential predictor $X$ and assume the true mean is given by
$E(Y \mid X) = \beta X$. The observed values $X_i$, $i = 1, \ldots, n = 11$, are the
same as the values of $X_{i1}$ in Example~1: $X_i = i - 1 - (n-1)/2$. Now, however,
we assume that $Y_i$ is observed without error for $i = 1, \ldots, n-1$, so that
$Y_i = \beta X_i$, and we assume that $Y_n = \beta X_n - K$. In this example, the first $n - 1$ training 
data points are observed without error because for now we are interested exclusively in 
quantifying the impact of the influential observation. After developing intuition in this
stylized example, we consider the usual setting where data are observed with error
in Sections~\ref{sec:meanshift} and \ref{sec:varianceinflation} and in the rest of the work.
The data are plotted in 
Figure~\ref{fig:slr_contaminants} for three different value of $K$.

There are two models under consideration: the null model $\Mn$ with only an intercept term,
and the full model $\Mone$ that includes $X$ as a regressor. Letting $\bX$ be the $n \times 1$ matrix of 
regressor values, under $L_2$ loss the Bayes estimators of the mean functions at $\bX$ for these two
models under Zellner's $g$ prior (\ref{eq:gprior}) are
\begin{eqnarray*}
	\hat{E}(\bY \mid \bX, \Mn) &=& \bbarY = -\bone_n \frac{K}{n}, \\
	\hat{E}(\bY \mid \bX, \Mone) &=& \bbarY + \frac{g}{1+g} \bX(\bX^T\bX)^{-1} \bX^T \bY \\
		&=& -\bone_n \frac{K}{n} + \frac{g}{1+g}\bX \bols.
\end{eqnarray*}

Suppose we observe testing data at $n$ design points $\tilde{X}_i$.
We assume the $\tilde{Y}_i$ values in the testing data follow the true regression line, 
$E(\tilde{Y}_i \mid \tilde{X}_i) = \beta \tilde{X}_i$,
plus independent Gaussian error~$\epsilon_i$ with mean zero and variance $\sigma^2$. Case $n$ in 
the test data set is assumed to be observed \emph{without} contamination $(K = 0)$. 
The $\mbox{Beta-Binomial}(1,1)$ prior over the model space assigns prior probabilities of $1/2$ to each
model and leads to the model-averaged predictions
\begin{eqnarray*}
	\hat{\bY} = \frac{g}{1+g} w(g, \bY) \tilde{\bX}\bols - \bone_n \frac{K}{n}, \label{eq:modavgpred}
\end{eqnarray*}
where $w(g, \bY) = \pi(\Mone \mid \bY)$ is the weight (posterior probability) for model $\Mone$.
We further define $s(g , \bY) = (g/(1+g))w(g, \bY)$ to be the \emph{shrinkage factor} for the model-averaged
estimate of $\beta$: the factor by which the least squares estimate of $\beta$ is shrunk toward zero when averaging
over the two models. The shrinkage factor $s(g,  \bY)$ is a non-linear function of $g$.

With the model-averaged predictions in hand, we can quantify predictive accuracy by computing the expected 
mean squared prediction error over the distribution of the test data $\tilde{\bY}$ as a function of $g$, conditioned 
on the observed data $\bY$:
\begin{multline*}
	E(MSPE(g) \mid \bY) = E\left[ \frac{1}{n} \sum_{i=1}^n (\tilde{Y}_i - \hat{Y}_i)^2 \right] \\
	= \frac{1}{n}\left[\left(\beta - s(g, \bY)\bols\right)^2 \left(\sum_{i=1}^n \tilde{X}_i^2\right) +
		2 \, \frac{K}{n} \left(\beta - s(g, \bY)\bols\right) \left(\sum_{i=1}^n \tilde{X}_i\right) \right]
			+ \frac{K^2}{n^2} + \sigma^2.
\end{multline*}
The expected mean squared prediction error is therefore minimized, as a function of $g$, when
\begin{equation}\label{eq:generalized_parallel_condition}
s(g, \bY) \bols = \beta + \frac{\sum_{i=1}^n \tilde{X}_i (K/n)}{\sum_{i=1}^n \tilde{X}_i^2},
\end{equation}
 i.e., when the model-averaged prediction line is parallel to 
 a line whose slope is the sum of two terms: the slope of the true regression line and the estimated least squares slope for a data set 
 in which the observations taken at the $n$ design points 
 $\tilde{X}_i$ are all equal to $K/n$.  
 
The optimal value of $g$ is thus seen to depend on the the true model slope, the testing design points, and the size $K$ of the contamination.  In a situation where the validation design points and the contamination size are random, one could make use of 
existing knowledge of certaing aspects of their distributions to obtain a plausible value for the second term in the RHS of Equation~(\ref{eq:generalized_parallel_condition}).  For example, expected values for $\sum_{i=1}^n \tilde{X}_i$, $\sum_{i=1}^n \tilde{X}_i^2,$ and $K$ could be substituted into the expression.  Of special interest is the situation where the testing design points $\tilde{X}_i$ coincide with the design points in the training data set.  If that happens, then $\sum_{i=1}^n \tilde{X}_i = 0$ and the 
optimal value of $g$ is the one that makes the model-averaged prediction line parallel to the true regression line. We focus on this situation when developing our methods.
The simulation examples in Section~\ref{sec:sims} examine the behavior of our method when the testing locations $\tilde{\bX}$ are allowed to differ from the training
locations $\bX$.

\begin{figure}
\centering
         \includegraphics[width=.8\linewidth]{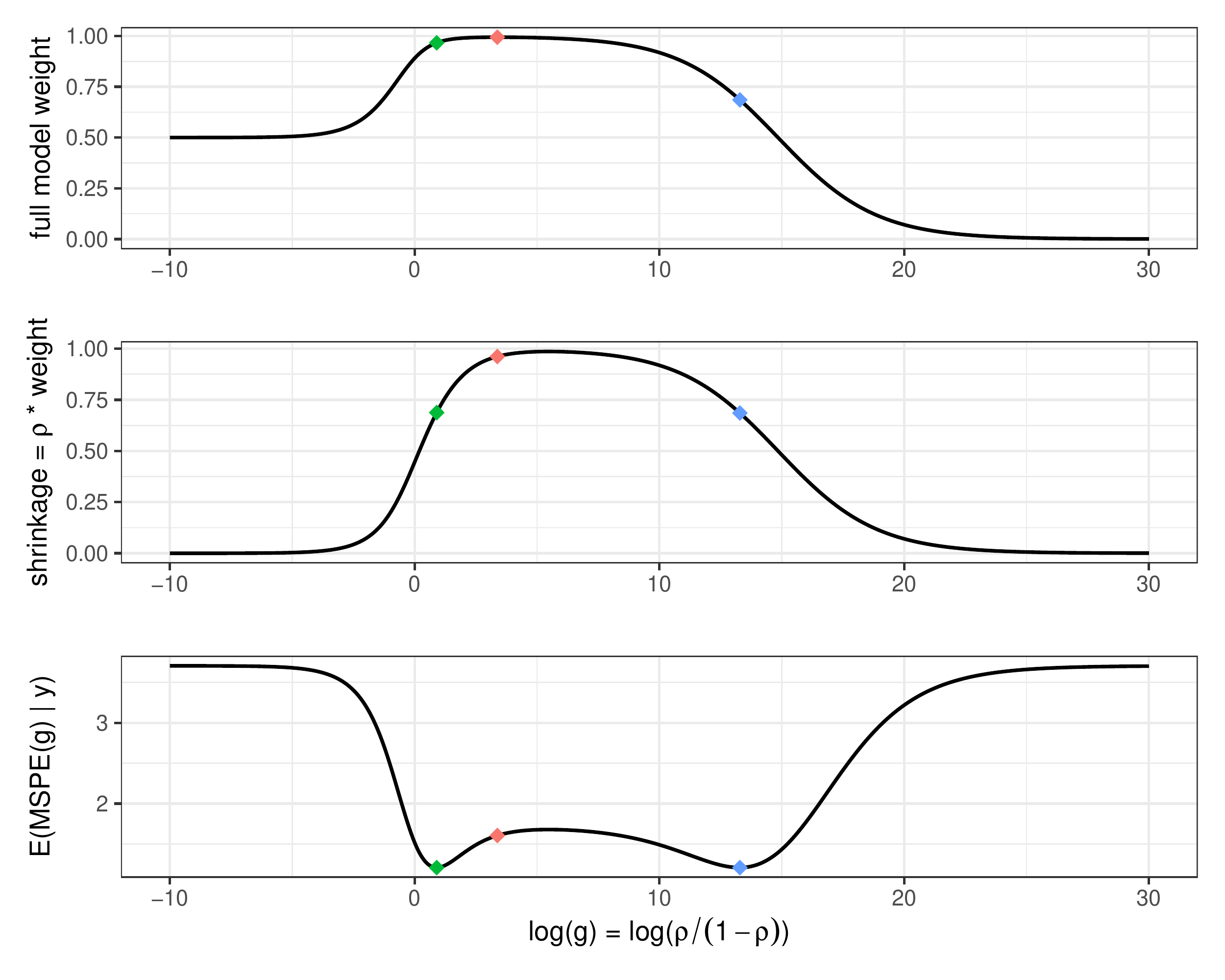}\\ 
         \caption{Behavior of $w(g, \bY)$ (top panel), $s(g, \bY) = \rho w(g, \bY)$ (middle panel) and
         $E(MSPE(g) \mid \bY)$ (bottom panel) as a function of $\log g$. The red diamonds correspond
         to the local empirical Bayes estimate of $g$; the green and blue diamonds correspond to the
         values of $g$ that minimize expected mean squared prediction error.}\label{fig:parallel_cond}  
\end{figure}

Having characterized the $g$ that leads to optimal predictions, we examine as a function of $g$
the behavior of the quantities $w(g, \bY)$, $s(g, \bY)$, and $E(MSPE(g) \mid \bY)$
when $n = 11$, $\beta = 0.5$, $K = -5$, and $\sigma^2 = 1$. These quantities are plotted in Figure~\ref{fig:parallel_cond}.
A question of interest is how the value of $g$ that minimizes expected mean squared prediction error described above compares
to other choices of $g$, e.g., the local empirical Bayes estimate. For interpretability, it is helpful to work on the 
transformed scale $\rho = g/(1 + g)$. The red diamond in the top panel of Figure~\ref{fig:parallel_cond} shows that the local empirical Bayes 
value of $\rho$, i.e., the value of $\rho$ that maximizes the posterior model weight, is given by $\rho = 0.967$ (or $\log g = 3.378$), 
yielding $w(g = e^{3.378}, \bY) = 0.994$. The red diamond in the middle panel tracks the corresponding shrinkage $s(g = e^{3.378}, \bY) = 0.962$, and the 
red diamond in the bottom panel tracks the corresponding expected mean squared prediction error, $E(MSPE(g = e^{3.378} \mid \bY)) = 1.604$.

The green diamond in the bottom panel, however, shows that the expected squared error of prediction is minimized for $\rho = 0.711$ (or $\log g = 0.901$) 
where $E(MSPE(g = e^{0.901} \mid \bY)) = 1.207$. The corresponding values of the model weight and shrinkage, tracked by the green diamonds in the top and 
middle panel, are $0.967$ and $0.688$. The optimal value of $\rho$ is much smaller than the one suggested by local empirical Bayes and much more shrinkage 
is needed to minimize the expected MSPE.

Figure~\ref{fig:parallel_cond} shows that additional interesting features occur for values of $\rho$ approaching 1 from the left ($g$ going to infinity). 
For one thing, as a consequence of Bartlett's paradox \citep{lian:08}, the posterior weight assigned to the regression model goes to zero and the null model 
gets fully weighted. Equation~(\ref{eq:modavgpred}) says that setting $\rho = 0$ ($g = 0$) and $\rho = 1$ ($g = \infty$) will yield the same predictions. 
Hence, the expected MSPE is the same at $\rho = 0$ and $\rho = 1$. As we move from the optimal value of $\rho$ (green diamond) toward $\rho$ equal to its 
local empirical Bayes value (red diamond), the predictive performance deteriorates. In this example, the local empirical Bayes performance is still better than 
the weak predictive performance for the extreme values $\rho = 0$ and $\rho = 1$. However, the presence of more influential cases can make the local 
empirical Bayes performance deteriorate even further and become even more similar to the performance that would be attained by ignoring the independent 
variable and predicting the mean response observed in the training data set.

Another interesting finding, tracked by the blue diamonds, is that the optimal expected predictive performance attained by setting $\rho = 0.711$ 
(or $\log g = 0.901$) is also attained by setting $\rho = 0.9999983$ (or $\log g = 13.288$) (geometrically, there are two ways to make the prediction and 
true regression lines parallel). However, one can show that there is a linear relation between the variance and the expectation of the MSPE, and that there is 
considerable instability in the predictive performance attained for $\rho$ values in a small neighborhood of $\rho = 0.9999983$. Confronting this with the 
considerable stability of the predictive performance attained for $\rho$ values in a sizable neighborhood of $\rho = 0.711$ suggests that setting empirically 
$\rho = 0.711$ (i.e., $\log g = 0.901$) in the prior specification would be a smart choice in this problem. Note that, in view of the role that $\rho$ plays in 
determining the shrinkage, it is most natural to seek stability on the $\rho$ scale rather than the $g$ or $\log g$ scale.

\begin{figure}
\centering
         \includegraphics[width=1\linewidth]{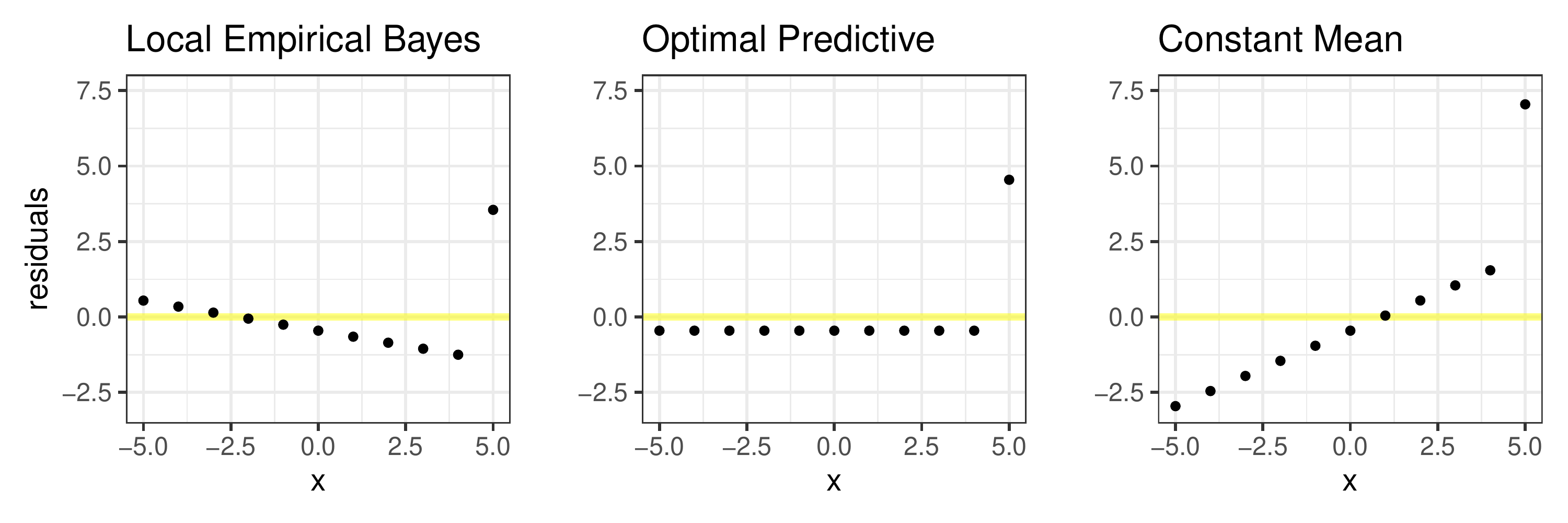}\\ 
         \caption{Residuals under three choices of $g$: $g = e^{3.378}$ (left panel),
         $g = e^{0.901}$ (center panel) and $g = 0 (\infty)$ (right panel).} \label{fig:parallel_resids} 
\end{figure}

We gain insight into these findings by analyzing, in Figure~\ref{fig:parallel_resids}, the behavior of the training data residuals, $\hat{Y}_i - Y_i$, 
corresponding to the various values of $\rho$ under consideration. The left panel corresponds to the residuals for the local empirical Bayes value. 
Aside from the residual for the influential observation, they exhibit a pattern which (while attenuated) mirrors that of the residuals for the constant prediction 
at the observed mean value plotted in the right panel. Although biased, the residuals corresponding to the optimal $\rho = 0.711$, displayed in the middle 
panel, do not show any pattern. This is in agreement with the previous finding that the expected MSPE is minimized when the prediction line is parallel to 
the true regression line and supports the intuition that good predictive performance and well behaved residuals go hand in hand.

\subsection{General Mean-shift Contamination}\label{sec:meanshift}
Section~\ref{sec:ex2} demonstrated, in a stylized, one-dimensional example, how $g$ could be chosen to minimize prediction error
when the training data had been contaminated by a single, influential mean-shift outlier. This section generalizes the result to
the setting where multiple candidate predictors exist, the data are contaminated according to the mean-shift contamination
described by \citet{abraham1978linear}, and the response variable is assumed to be observed with error. Suppose the data are generated 
from the model
\begin{equation}\label{eq:model1}
  Y_i=\begin{cases}
    \alpha + \bm{x}_i^T\bm{\beta}_{True}+\epsilon_i+K, & \text{w.p. $\pi$},\\
    \alpha + \bm{x}_i^T\bm{\beta}_{True}+\epsilon_i, & \text{w.p.  $1-\pi$},
  \end{cases}
\end{equation}
where $\epsilon_i \stackrel{iid}{\sim} N(0,\sigma^2)$ and $\bb_{True}$ is a $p$-dimensional vector which may contain zeros.
Let $\bX$ be the $n\times p$ matrix with rows $\bm{x}_i^T$ and column means zero, and let $\bY$ be the response vector.
Under this model, each case is contaminated independently of the others with probability~$\pi$. For simplicity, in this section we 
assume $\alpha = 0$; this assumption does not affect the conclusions we draw. 

Assume that for each model $\Mg$ we assign Zellner's $g$ prior to the regression coefficients, $\bbg$, and that we allow the 
scale parameter, $\gg$, to differ across models. Denote the collection of scale parameters as $\bg$. For a given prior over the
model space, denote the model-averaged estimate of the vector of regression coefficients by $\hbbbma(\bg, \bm{X}, \bm{Y})$.
With the understanding that this estimate depends on $\bm{g}$, we will refer to this estimate as $\hbbbma$ for short. Under this model,
the vector of model-averaged fitted values is $\hat{\bY} = \bbarY + \bX\hbbbma$.

Suppose future observations $\tilde{\bY}$ are generated from model (\ref{eq:model1}) using the same design matrix $\bX$ and a
possibly different contamination proportion $\tilde{\pi}$. We can examine
how the choice of the $\gg$ is connected to predictive accuracy by examining the expected sum of squared prediction errors:
\begin{eqnarray}
	E(SSPE(\bg) \mid \bY, \bX) &=& E(( \tilde{\bY} - \hat{\bY})^T( \tilde{\bY} - \hat{\bY}) \mid \bX, \bY, \bg) \nonumber \\
		&=& (\hat{\bY} - E(\tilde{\bY}))^T(\hat{\bY}-E(\tilde{\bY})) + \nonumber \\
		&& \hspace{0.5in}
			E((\tilde{\bY}-E(\tilde{\bY}))^T(\tilde{\bY}-E(\tilde{\bY})) \mid \bX, \bY, \bg). \label{eq:sspe1}
\end{eqnarray}
The first term in the final expression for the LHS of (\ref{eq:sspe1}) can be viewed as a squared bias term, while the 
second term can be viewed as a variance term. The variance
term can be shown to be $n\sigma^2$, which does not depend on the $\gg$, while the squared bias term is
\[
	(\hat{\bY}-E(\tilde{\bY}))^T(\hat{\bY}-E(\tilde{\bY})) =
		(\bX\hbbbma - \bX \bb_{True})^T(\bX\hbbbma - \bX\bb_{True}) +
			n(\bar{Y}-K\tilde{\pi})^2.
\]
The term $n(\bar{Y} - K\tilde{\pi})^2$ doesn't depend on the
values in $\bm{g}$. If values of $\bg$ exist such that $\hbbbma = \bb_{True}$, those values of $\bg$ would
minimize the prediction error. As in the simple case in Section~\ref{sec:ex2}, a model-averaged regression plane that is parallel to the true mean function will
result in optimal predictions as measured by expected $SSPE(\bg \mid \bY, \bX)$.

\subsection{Variance-inflation Contamination}\label{sec:varianceinflation}
Suppose the data are generated from the variance-inflation contamination model of \cite{box1968bayesian},
\begin{equation}\label{model2}
  Y_i=\begin{cases}
    \alpha + \bm{x}_i^T\bm{\beta}_{True}+\sqrt{K}\epsilon_i, & \text{w.p. $\pi$},\\
    \alpha + \bm{x}_i^T\bm{\beta}_{True}+\epsilon_i, & \text{w.p.  $1-\pi$},
  \end{cases}
\end{equation}
where $\epsilon_i \stackrel{iid}{\sim} N(0,\sigma^2)$. Again, $\bb_{True}$ is the $p$-dimensional true regression parameter, 
which can contain zeros, and we assume $\alpha = 0$ to simplify the exposition.

As in Section~\ref{sec:meanshift}, denoting by $\tilde{\pi}$ the contamination proportion in the testing data,
$E(SSPE(\bg) \mid \bX, \bY)$ can be decomposed into a squared bias term and a variance term. 
The variance term, $n\sigma^2(K\tilde{\pi} + 1 - \tilde{\pi})$, does not depend on $\bg$, while the squared bias term is
\[
	(\bX \hbbbma - \bX \bb_{True})^T(\bX \hbbbma - \bX \bb_{True}) +
		n\bar{Y}^2.
\]
The conclusion is the same: the BMA regression plane that is parallel to the true regression plane will minimize the expected sum
of squared prediction errors, if such a plane exists as a function of $\bg$.

\section{Proposed Methods}\label{sec:methods}
BMA predictions in regression depend on the posterior distribution over models and on the intra-model estimates of the regression
coefficients, all of which depend on the values of $g_\gamma$ in Zellner's $g$ prior. These quantities are sensitive to influential outliers,
as demonstrated in the example in Section~\ref{sec:ex1}. Section~\ref{sec:ex2} established, in a simple example, that in the presence
of influential outliers prediction error can be minimized if $g$ can be chosen to make the model-averaged regression plane parallel to
the true regression plane. Choosing $g$ in this way has the effect of producing well-behaved residuals, which we know from classical 
regression analysis is desirable. Sections~\ref{sec:meanshift} and Section~\ref{sec:varianceinflation} showed that this intuition holds for 
more complex model settings.

The intuition developed in Section~\ref{sec:intuition} provides motivation for how one might think about prior specification when there is concern that
outliers might impact several (or many) of the models in the ensemble: choose values for the $g_\gamma$ to make the model-averaged 
regression plane parallel to the true regression plane, $\hbbbma = \bb_{True}$. We refer to this as the ``parallel condition.'' 
There are two obvious practical limitations to this approach. The first is, of course, that the orientation of the true regression plane is defined by 
$\bb_{True}$, the unknown parameters of the model. The second is that, even if the true $\bb_{True}$ were known, there is no guarantee that 
values $g_\gamma$ exist that produce model-averaged coefficients satisfying the parallel condition. This can be illustrated simply in the context 
of the example in Section~\ref{sec:ex2} by taking $K = 5$ instead of $K = -5$. When $K = 5$, the least squares estimate of the slope is 
$\bols = 0.2727$. As a function of $g$, the shrinkage coefficient $s(g, \bY) = (g/(1+g))w(g, \bY)$ ranges between zero and one, and so there is no 
value of $g$ that produces $\hat{\beta}_{BMA} = s(g, \bY)\bols$ equal to $\beta = 0.5$. 
This extends to the general setting where $p>1$.
In this case the model-averaged estimate of the coefficient corresponding to regressor $X_j$ is
\[
	\hat{\beta}_{BMA,j} = \sum_{\gamma \in \Gamma \; : \; X_j \in \Mg} \left(\frac{\gg}{1+\gg}\right) \hat{\beta}_{LS,\gamma, j} \pi(\Mg \mid \bY),
\]
where $\hat{\beta}_{LS,\gamma, j}$ is the least squares estimate of $\beta_j$ in a model $\Mg$ that contains $X_j$ as a regressor.
The terms $\hat{\beta}_{LS,\gamma, j}$ do not depend on the $\gg$ and can be bounded above and below by their largest and smallest values 
across the model space. The terms $(\gg/(1+\gg))\pi(\Mg \mid \bY)$ depend on the $\gg$ but are bounded below and above by zero and one, hence there 
is no guarantee that there exist values of $\gg$ that will produce a regression plane that is parallel to the true regression plane with 
$\hat{\beta}_{BMA,j} = \hat{\beta}_{True, j}$, for $j = 1, \ldots, p$.

In this section we address these practical limitations in several ways. First, in Section~\ref{sec:nonzerog}, we expand the space of prior distributions 
so that the parallel condition is achievable for a larger space of observed data sets than is possible under Zellner's $g$ prior
that shrinks toward zero. Second, in Section~\ref{sec:relax} we relax the strict parallel condition by attempting to find values of the model's
hyperparameters that make the model-averaged regression plane as close to parallel to the true regression plane
as possible in $L_2$ distance. Finally, in Section~\ref{sec:nullmix} we synthesize these ideas and propose an approach for choosing empirically the values of the 
model's hyperparameters with the goal of yielding small prediction error in the presence of influential outliers. As the orientation of the
true regression plane is unknown, we propose using robust estimates that are insensitive to influential outliers as part of the procedure
for choosing the hyperparameters.

\subsection{Expanding the Prior Model}\label{sec:nonzerog}

For particular data and true model settings there may not exist values $\gg$ under Zellner's $g$ prior (\ref{eq:gprior}) which shrinks all coefficients toward 
zero that satisfy $\hbbbma = \bb_{True}$. When this is the case, we might instead chose values $\gg$ that
make the model-averaged regression plane as close to parallel as possible to the true regression plane. To improve the quality of this approximation,
we expand the prior model to allow for shrinkage toward a potentially non-zero target, $\btg$. Under this prior, described in (\ref{eq:nzgprior}), the 
model-averaged estimate of the coefficient corresponding to regressor $X_j$ is
\[
	\hat{\beta}_{BMA,j} = \sum_{\gamma \in \Gamma \; : \; X_j \in \Mg} 
		\left(\frac{1}{1 + \gg}\theta_{\gamma, j} + \frac{\gg}{1+\gg} \hat{\beta}_{LS, \gamma, j}\right) \pi(\Mg \mid \bY).
\]
The hyperparameters $\btg$ provide extra flexibility that allows for 
shrinkage toward targets other than zero.
To set notation, let $\{\btg, \gg\}$ denote the hyperparameters for model $\Mg$ and let $\{\bTh, \bg\}$ denote the collection of hyperparameters $\btg$ 
and $\gg$ across all models indexed by $\gamma \in \Gamma$.
By enlarging the collection of hyperparameters from $\bg$ to $\{\bTh, \bg\}$, we are able to achieve
the parallel condition, $\hbbbma = \bb_{True}$, for a wider range of data and true model settings.

This can be seen clearly in a modified version of the example in Section~\ref{sec:ex2} where $K=5$ and the true regression coefficient
is $\beta = 2$. When shrinking toward a prior mean of zero, no value of $g$ is able to produce $\hat{\beta}_{BMA} = \beta$ because
$\bols = 1.7727$ and $\beta = 2 > \bols$. However, when shrinking toward a potentially non-zero prior mean $\theta$, there are many such
$\{\theta, g\}$ pairs that produce $\hat{\beta}_{BMA} = 2$, and we can chose a pair according to some rule (e.g., favoring small values of $\theta$
to encourage shrinkage toward zero, or discouraging small values of $g$ to avoid overconfidence in the prior). The added flexibility of shrinking
toward a non-zero mean $\theta$ does not guarantee we can achieve the parallel condition, e.g., when $\beta = 0.5$ and $K = 5$ in the example
in Section~\ref{sec:ex2}, there are no $\{\theta, g\}$ pairs that produce $\hat{\beta}_{BMA} = 0.5$. The added flexibility, though, means we can do
no worse than shrinking toward zero if our goal is to bring the model-averaged prediction plane as close to parallel to the true regression plane
as possible, and so we work with prior (\ref{eq:nzgprior}) from now on.

\subsection{Relaxing the Parallel Condition: The Overall-Mixture Prior
Specification}\label{sec:relax}
Even after expanding the space of priors to include non-zero prior means $\bm{\theta}_\gamma$, the parallel condition is not guaranteed to
be achievable for all data sets. To address this issue, we might relax the parallel condition and choose hyperparameter 
values $\{\bTh, \bg\}$ so that the model-averaged regression plane is as close in $L_2$ distance as possible to the true regression
plane:
\begin{equation}
	\{\hat{\bTh}, \hat{\bg}\} = \underset{\bTh, \bg}{\arg\min} \,
		|\!| \hbbbma - \bb_{True} |\!|^2. \label{eq:L2relax}
\end{equation}
This defines an empirical Bayes approach for selecting hyperparameter values that could be implemented if $\bm{\beta}_{True}$ were known
or replaced with a suitable estimate.
We call this approach the {\em overall-mixture\/} prior specification.
The main difficulty in implementing (\ref{eq:L2relax}) in practice is computational: with $p$ predictors,
there are $(2^p - 1) + (p2^{p-1})$ total hyperparameters: $2^p-1$ scale parameters $g_\gamma$ and $\sum_{k=0}^p k {p\choose k} = p2^{p-1}$
individual location parameters $\theta_{\gamma, k}$, where $\theta_{\gamma, k}$ is the $k$th elements of the prior mean vector for model $\Mg$.
Not only does this represent a high-dimensional optimization problem, but for
any given collection of values $\{\bTh, \bg\}$ calculating $\hbbbma$ requires summing over
$2^p$ models, where each term in the sum contains non-linear functions of elements of $\{\btg, \gg\}$.
Brute-force, numerical optimization approaches might be feasible in
very small problems, but will be challenging in general.
If interest is limited to point prediction under posterior predictive $L_2$ loss
and the 
the parallel condition can be achieved, 
predictions based on the resulting $\hbbbma$
would coincide with predictions based 
directly on 
${\bm{\beta}}_{True}$.

\subsection{Local Null-Mixture Prior Specification}\label{sec:nullmix}
To ease the computational burdens discussed in Section~\ref{sec:relax} while still using the parallel condition to our advantage, we propose
a new local empirical Bayes approach for hyperparameter specification. The approach is ``local'' in that each model receives its own, unique
hyperparameter values $\{\btg, \gg\}$ and also in that the selection of values for the hyperparameters requires
calculations that are dependent \emph{only} on that model. In principle, this is similar to the local empirical Bayes approach,
though our criterion for specifying the hyperparameter values is motivated by the parallel condition (rather than by maximization of the marginal
likelihood).

Rather than focusing on the full model-averaged regression plane defined by $\hbbbma$ in~(\ref{eq:L2relax}), which requires
knowledge about all $2^p$ models in $\Gamma$, for any given model $\Mg$ we focus solely on the relationship between the model $\Mg$
and the null model with no predictors, $\Mn$. If the only two models under consideration were $\Mg$ and $\Mn$, the posterior probability
of model $\Mg$ would be
\[
	\pi^*(\Mg \mid \bY) = \left(1 + \frac{m(\bY \mid \Mn)}{m(\bY \mid \Mg)}\frac{\pi^*(\Mn)}{\pi^*(\Mg)}\right)^{-1} =
	\left(1 + \frac{1}{BF(\Mg  :  \Mn)}\frac{\pi^*(\Mn)}{\pi^*(\Mg)}\right)^{-1},
\]
where $BF(\Mg : \Mn)$ is the Bayes factor for comparing model $\Mg$ to the null model given in Equation~(\ref{eq:nzbf}).
We use $\pi^*(\Mn)$ and $\pi^*(\Mg)$ to denote the assigned prior model probabilities when the model space contains only the
two models $\Mn$ and $\Mg$. As a default, we use $\pi^*(\Mn) = \pi^*(\Mg) = 0.5$, which is also what would result from using the 
uniform prior or the Beta-Binomial$(1,1)$ 
prior over a model space that contains only these two models.
The model-averaged estimate of $\bm{\beta}_\gamma$ when the only two models in the model space are 
$\Mn$ and $\Mg$ is
\[
	\hbb_{BMA, \gamma} = \pi^*(\Mg \mid \bY) \left(\frac{1}{1+\gg}\bm{\theta}_\gamma + \frac{\gg}{1+\gg} \hbb_{LS, \gamma} \right),
\]
where $\hbb_{LS, \gamma}$ is the
least squares estimate of $\bbg$. The relaxed parallel condition in (\ref{eq:L2relax}) then tells us to choose hyperparameter values that 
satisfy
\begin{equation}
	\{ \hat{\bt}_\gamma, \hat{g}_\gamma\} = \underset{\{\btg, \gg\}}{\arg\min} \,
		|\!| \hbb_{BMA, \gamma} - \bb_{True, \gamma} |\!|^2,
		\label{eq:L2nullmix}
\end{equation}
where $\bb_{True, \gamma}$ is interpreted here as the true regression coefficients for model $\Mg$. Once hyperparameters for each of the 
$2^p-1$ models have been found via (\ref{eq:L2nullmix}), model-averaged prediction proceeds as usual using
the posterior model probabilities $\pi(\Mg \mid \bY)$ computed under the \emph{original} prior $\pi(\Mg)$ over the \emph{entire}, 
unrestricted model space, $\Gamma$. 

The local null-mixture approach for selecting hyperparameter values defined in (\ref{eq:L2nullmix}) attempts to make the locally-model-averaged
regression plane parallel to an unknown regression plane oriented according to $\bm{\beta}_{True, \gamma}$. In practice, we require
estimates of these unknown parameters. We opt to use an estimate of $\bm{\beta}_{True, \gamma}$ that is robust in the sense that is 
relatively insensitive to influential outliers \citep[see][for a general treatment of robust regression]{rous:05}. For a given model $\Mg$,
we rank each observation according to its Cook's distance, $D_{i,\gamma}$ \citep{Cook:dete:1977}. We then remove 10\% of the cases
corresponding to the largest values of Cook's distance, and compute the ordinary least squares estimate of the regression coefficients using
the remaining 90\% of the cases. With this robust estimate, $\hat{\bm{\beta}}_{Robust, \gamma}$, in hand, we choose hyperparameters
for model $\Mg$ that satisfy the criterion
\[
	\{ \hat{\bt}_\gamma, \hat{g}_\gamma \} = \underset{\{ \btg, \gg \}}{\arg\min} \,
		|\!| \hbb_{BMA, \gamma} - \hat{\bm{\beta}}_{Robust, \gamma} |\!|^2_2.
\]
We use the \texttt{optim} function in R \citep{R} to implement this approach in the examples in Section~\ref{sec:sims}.

The local null-mixture approach is attractive due to the fact that it
reduces the computational burden associated with criterion
(\ref{eq:L2relax}).  However, because the hyperparameters are chosen
locally---by mixing models $\Mg$ and $\Mn$---rather than globally---by
averaging over all models in (\ref{eq:L2relax})---the predictive
optimally associated with the parallel condition is not
guaranteed. However, based on the intuition developed in
Section~\ref{sec:ex2}, choosing hyperparameters locally will still
result in well-behaved residuals that are local to the mixture of the
models $\Mg$ and $\Mn$. As good prediction and well-behaved residuals
tend to go hand-in-hand, we expect the local null-mix method to
perform well in practice. In addition, as the simulations in Section~\ref{sec:sims} will
demonstrate, this local approach can outperform the global approach when some
of the true regression parameters $\beta_j$ are equal to zero.

\section{Simulations}\label{sec:sims}

In this section we present simulation studies comparing the
predictive performance of our proposed method to that of the related
methods mentioned in Section~\ref{sec:regression_model}: local
empirical Bayes (EB-Local), global empirical Bayes (EB-Global), and
the hyper-$g/n$ methods. The computations for the related methods were
performed using the \texttt{BAS} package in \texttt{R} \citep{BAS}. 
The $\mbox{Beta-Binomial}(1,1)$ prior over the model space with 
 $\pi(\Mg) = (p+1)^{-1} {p \choose |\Mg|}^{-1}$ was used for each of the 
BMA methods. We also compare the performance of the BMA predictions to
the performance of predictions based on the robust estimate $\hat{\bm{\beta}}_{Robust, \gamma}$
described in Section~\ref{sec:nullmix} under the full model with 
$\gamma = (1, \ldots, 1)^T$ (i.e., no model averaging).

In real-world settings, data contamination may constitute a one-off
occurrence or it may represent a structural component of the
underlying stochastic mechanism (a fraction of the observations may
follow a different distribution than the bulk of the observations). In
the first case we might expect only the training (or only the testing)
data to be contaminated.  In the second case we would expect both the
training and the testing data to be contaminated.  Our simulations
evaluate the performance of the different methods in both of these
cases, as well as in the case when neither the training nor the testing
data are contaminated.  We consider contaminations arising from a
mean-shift and from a variance-inflation scheme.  The data
contamination patterns for the five simulations we conducted are
summarized in Table~\ref{tab:contamination_scheme}.

\begin{table}[t] \centering
  \begin{tabular}{lccccc} \cline{2-6} &
\multicolumn{5}{c}{contamination pattern} \\ \cline{2-6} 
 & M-S/no & M-S/M-S & V-I/no & V-I/V-I & no/no \\ \hline \hline
training data & M-S & M-S & V-I & V-I & no \\
\hline testing data & no & M-S & no & V-I & no
\\ \hline \hline
  \end{tabular}
  \caption{Data contamination patterns for the training and testing data
    used in the simulations:
    mean-shift (M-S), variance-inflation (V-I), and no contamination (no).}\label{tab:contamination_scheme}
\end{table}

\begin{table}[t] \centering
  \begin{tabular}{lccccccccc} 
\cline{2-10} 
& \multicolumn{5}{c}{$p=5$} & \hspace{0.5cm} &
\multicolumn{3}{c}{$p=10$} \\
& \multicolumn{5}{c}{complexity level} & &
\multicolumn{3}{c}{complexity level}\\
\cline{2-10}  
 & 1 & 2 & 3 & 4 & 5 & & 1 & \hspace{0.45cm} 5 \hspace{0.45cm} & 10 \\ \hline \hline
$\beta_1 = $        & \,1\, & \,1\,& \,1\, & \,1\, & \,1\, & & \,0.5\, & \,0.5\, & \,0.5\, \\
\hline $\beta_2 = $ & 0 & 2 & 2 & 2 & 2 &  & 0.0 & 1.0 & 1.0 \\
\hline $\beta_3 = $ & 0 & 0 & 3 & 3 & 3 & & 0.0 & 1.5 & 1.5 \\
\hline $\beta_4 = $ & 0 & 0 & 0 & 4 & 4 & & 0.0 & 2.0 & 2.0\\
\hline $\beta_5 = $ & 0 & 0 & 0 & 0 & 5 & & 0.0 & 2.5 & 2.5 \\ 
\hline $\beta_6 = $ &  &  &  &  &  & & 0.0 & 0.0 & 3.0 \\ 
\hline $\beta_7 = $ &  &  &  &  &  & & 0.0 & 0.0 & 3.5 \\ 
\hline $\beta_8 = $ &  &  &  &  &  & & 0.0 & 0.0 & 4.0 \\ 
\hline $\beta_9 = $ &  &  &  &  &  & & 0.0 & 0.0 & 4.5 \\ 
\hline $\beta_{10} = $ &  &  &  &  &  & & 0.0 & 0.0 & 5.0 \\ 
\hline \hline
  \end{tabular}
  \caption{ The columns in the left-hand side of the table display the values of 
    $\bm{\beta} = (\beta_1, \beta_2, \beta_3, \beta_4, \beta_5)^T$ for the five 
    models of increasing complexity used in the simulations with $p=5$.
    The columns in the right-hand side of the table display the values of 
    $\bm{\beta} = (\beta_1, \ldots, \beta_{10})^T$ for the three 
    models of increasing complexity used in the simulations
    with $p=10$.
    }\label{tab:complexity_level}
\end{table}

Prior to possibly being contaminated, the training and the testing data 
are both generated from the true model  
\begin{equation*}
\begin{split}
\bm{Y}_{Train}&=\bm{X}_{Train}\bm{\beta}+\alpha\bm{1}+\bm{\epsilon},\\
\bm{Y}_{Test}&=\bm{X}_{Test}\bm{\beta}+\alpha\bm{1}+\bm{\epsilon}.
\end{split}
\end{equation*}
For simplicity, in all simulations, we set $\alpha=0$.  The design
matrices $\bm{X}_{Train}$ and $\bm{X}_{Test}$ have dimension
$100 \times 5$ and their rows 
are independently generated from a multivariate normal distribution,
MVN$(\bm{0}, \bm{\Sigma})$, where the covariance matrix
$\bm{\Sigma}$ has diagonal elements equal to $1$ and off-diagonal
elements equal to $0.6$.
The error vector $\bm{\epsilon}$ has i.i.d.\ standard normal random
elements.

In the mean-shift case, we contaminate the data by randomly selecting
5\% of the observations and adding $K=10$ to the dependent variable.
In the variance-inflation case, we contaminate the data by randomly
selecting 5\% of the observations and multiplying their errors, $\epsilon_i$, by $\sqrt{10}$ before adding
them to the mean function so that the errors for the contaminated cases have variance $K = 10$.
To assess how model complexity affects
performance we consider five models of increasing complexity where the number of active
predictors ranges from one to five. The values of the vectors of regression
coefficients $\bm{\beta}$ for the five models are summarized in the left-hand side of
Table~\ref{tab:complexity_level}.

For settings where the training data are contaminated, we use either
the mean-shift scheme or the variance-inflation scheme to contaminate
the training data set. For the testing data, we either contaminate it
with the same scheme used to contaminate the training data set or we
leave it uncontaminated. Therefore, we have multiple settings based
on the true coefficients and the contamination scheme of the training
and testing data sets. Since we have 5 true $\bb$ 
vectors and 5 different contamination combinations, 
we have 25 different settings in total.

Our simulations consider all $5 \times 5 = 25$ settings resulting from
pairing one of the 5 data contamination patterns with one of the 5
model complexity levels.  In each of these settings, we simulate 
500 training and testing data sets to evaluate the performance of the
different methods mentioned at the start of the section.  
We apply each method to the training data sets to estimate models and let the
models make predictions $\bm{\hat{Y}}$ for the testing data
sets. Specifically, the BMA predictions $\bm{\hat{Y}}$ are calculated
according to the following steps.

\begin{enumerate}
\item \textbf{Centering the training data:} We use $\bm{X}_{Train,c}$ to denote the
  centered training design matrices, and $\bm{X}_{Train,mean}$ to
  denote the corresponding column mean vector,~i.e.,
\begin{equation*}
\bm{X}_{Train,c}=\bm{X}_{Train}-\bm{1} \, \bm{X}_{Train,mean}^T,
\end{equation*}
where $\bm{1}$ is a vector of size $n$ with all the elements equal to 1.
\item \textbf{Fitting models to the training data sets}: By
  applying each of the BMA methods, indexed by $m$, to the centered
  training data sets $(\bm{Y}_{Train},\bm{X}_{Train,c})$ we obtain
  fitted BMA planes:
\begin{equation*}
  \bm{\hat{Y}}_{Train,m}=\bm{1} \, \bar{\bm{Y}}_{Train}+\bm{X}_{Train,c}\,
  \bm{\hat{\beta}}_{BMA,m},
\end{equation*}
where $\bm{\hat{\beta}}_{BMA,m}$ is the BMA estimate
of $\bm{\beta}$ for method
$m$, $\bar{\bm{Y}}_{Train}$ is the average of $\bm{Y}_{Train}$ (which
is also the estimated intercept), and $\bm{\hat{Y}}_{Train,m}$ are the
values predicted by method~$m$ for the training data.
\item \textbf{Making predictions:} 
  We employ the estimated parameters
  from the previous step
  to construct the following prediction plane:
\begin{equation*}
  \hat{\bm{Y}}_{m}=\bm{1} \, \bar{\bm{Y}}_{Train}+\left(\bm{X}_{Test}-
    \bm{1} \, \bm{X}_{Train,mean}^T\right)\bm{\hat{\beta}}_{BMA,m}.
\end{equation*}
By subtracting the column means of the training design matrix
from the testing design matrix we ensure that the estimated model
parameters can be meaningfully applied to the testing data.
\end{enumerate}

In each of the 25 simulation settings and for each method~$m$,
after acquiring the predictions $\bm{\hat{Y}}_{m}^{(i)}$
for replication $i$, $i=1, \ldots, 500$,
we compute the observed MSPE for that iteration as
\[
\text{MSPE}_{m}^{(i)}=
\frac{1}{n}\left\|{\bm{\hat{Y}}}_{m}^{(i)}-
  \bm{Y}_{Test}^{(i)}\right\|^2,
\]
where $n$ is the number of observations in the testing data set (here
$n=100$).
We are mainly interested in 
assessing the relative performance of the various
methods. To this end, for replication $i$,
letting $m^*$ denote the hyper-$g/n$ method that we take
as a reference, we compute the relative percent reduction in MSPE for the
other methods relative to
$m^*$ as
\[
  \text{RR}_{m, m^*}^{(i)} = 100 \times
\frac{\text{MSPE}_{m^*}^{(i)} -
  \text{MSPE}_{m}^{(i)}}{\text{MSPE}_{m^*}^{(i)} }, \,\,\,
  i=1, \ldots, 500.
\]

\begin{figure}[t]
\begin{center}
\includegraphics[scale=0.5]{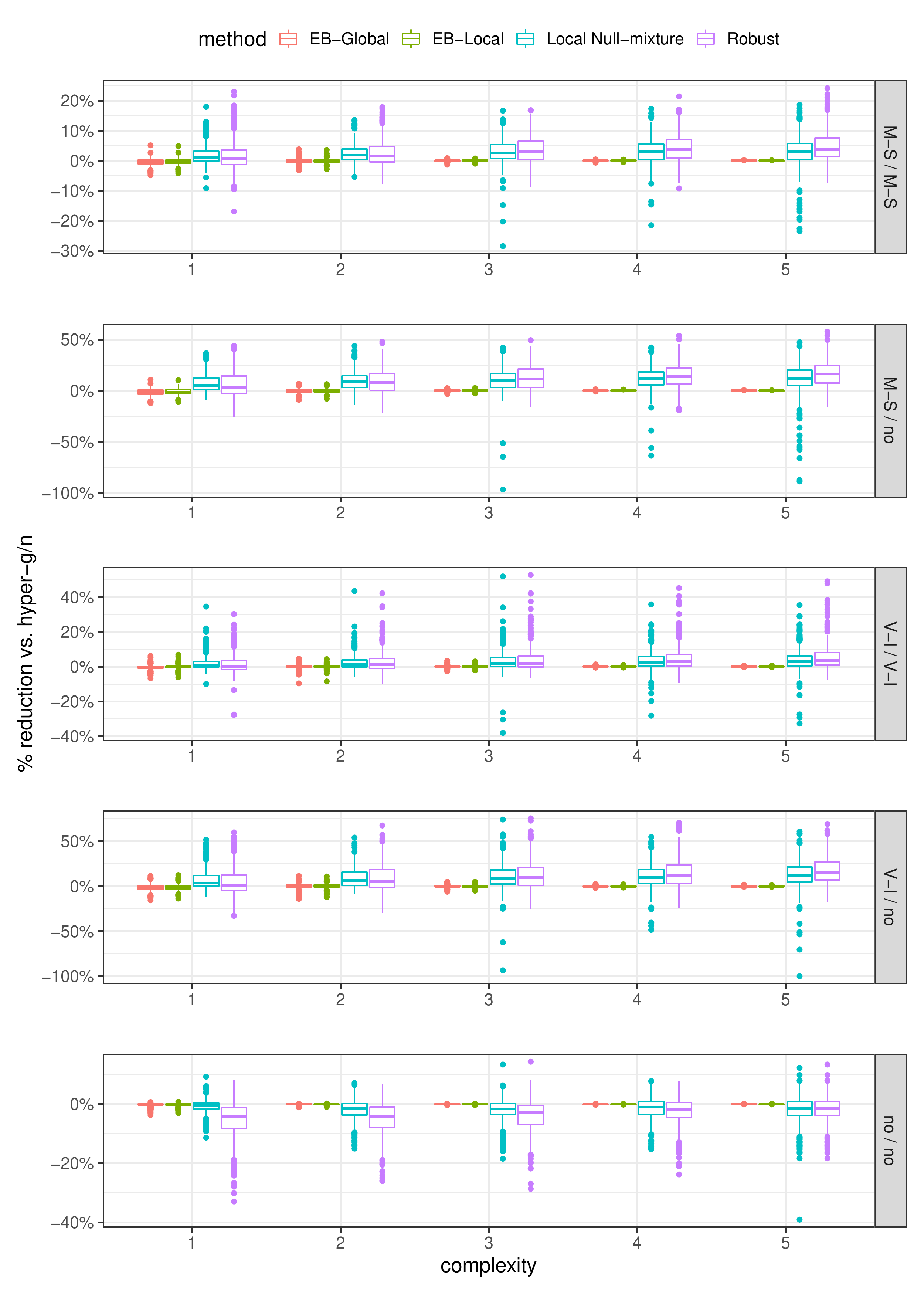}
\end{center}
\caption{Percent reduction in MSPE relative to
  the model-averaged hyper-$g/n$ method for the simulation study
  with $p=5$ predictors under
  complexity settings (the number of non-zero coefficients) ranging
  from 1 to 5. Each row corresponds to a contamination scheme,
  labeled ``X / Y'', where X (Y) denotes the contamination scheme for the
  training (test) data. ``M-S'' denotes mean-shift contamination, ``V-I''
  denotes variance-inflation contamination, and ``no'' denotes no
  contamination.
  }
\label{fig_MSPE_p5}
\end{figure}

The relative percent reduction in MSPE values from 500
replications for
each of the 25 simulation settings are summarized graphically in the
boxplots of Figure~\ref{fig_MSPE_p5}.  
First, we note that the two empirical Bayes BMA methods both
perform similarly to the hyper-$g/n$ BMA method. Comparing the
local null-mixture method to the other BMA methods,
perhaps the most conspicuous
finding is that, when the training data are contaminated and the
testing data are uncontaminated, the proposed method exhibits the
largest relative improvements over the 
competing BMA methods
for both contamination schemes (when making visual comparisons, 
note that the scale on the vertical
axes differ across panels).  The relative improvements are smaller in
settings when both the training and the testing data are
contaminated. The variability of the proposed method appears to
increase with model complexity.  In particular, when the complexity
level equals 5, the relative performance of the proposed method can be
significantly worse in a small number of replications.  The bottom
panel in the figure shows that the proposed methodology suffers a
little compared to the other BMA methods when the training data 
are uncontaminated. This behavior is to
be expected and suggests that the methodology is best suited for
situation in which the analyst suspects that contamination is present,
although the median performance deterioration in the case of no 
contamination is tolerable.

It follows from the last remark in Section~\ref{sec:relax}
that,
when the parallel condition can be achieved, 
predictions based directly on $\hat{\bm{\beta}}_{Robust, \gamma}$,
the robust estimate of ${\bm{\beta}}_{True}$ under the full model,
coincide with the predictions produced by the overall-mixture
method under posterior predictive $L_2$ loss. 
Compared to the local null-mixture method, the performance of the
robust (overall-mixture) predictions varies with the complexity of the
underlying, true, uncontaminated model.

When the true complexity is less than $p$, the performance of the
overall-mixture method suffers from its strong reliance 
on the robust estimate under an overparameterized model
and the attendant loss of accuracy from estimation of the $\beta_j$
coefficients equal to zero.
By contrast, the local null-mixture model attains higher precision
because it builds upon separate robust estimates
for all possible models including the model that excludes the $\beta_j$
equal to zero.  This is evidenced by the reduced variability 
exhibited by the boxplots (local null-mixture vs. robust) 
for smaller complexity settings in Figure~\ref{fig_MSPE_p5}.

While less variable overall, the EB-Global and EB-Local methods tend
to underperform relative to the local null-mixture and the robust
method whenever any contamination is present.
When neither the trainig nor the testing data are contaminated,
the EB-Global and EB-Local methods exhibit better predictive
performance.  In such cases, 
the local null-mixture method
outperforms the robust approach when the complexity is small
relative to $p$, and does about as well as the robust approach when
the complexity matches $p$.

In summary, these results suggest that, in the realistic situation
when there is uncertainty about model composition, the local
null-mixture method is preferable to the the robust method from a
predictive perspective and has the additional advantage of providing
posterior descriptions of uncertainty about variable inclusion and
other aspects of the posterior and predictive distributions.

\afterpage{\clearpage}

\begin{figure}[t]
\begin{center}
\includegraphics[scale=0.5]{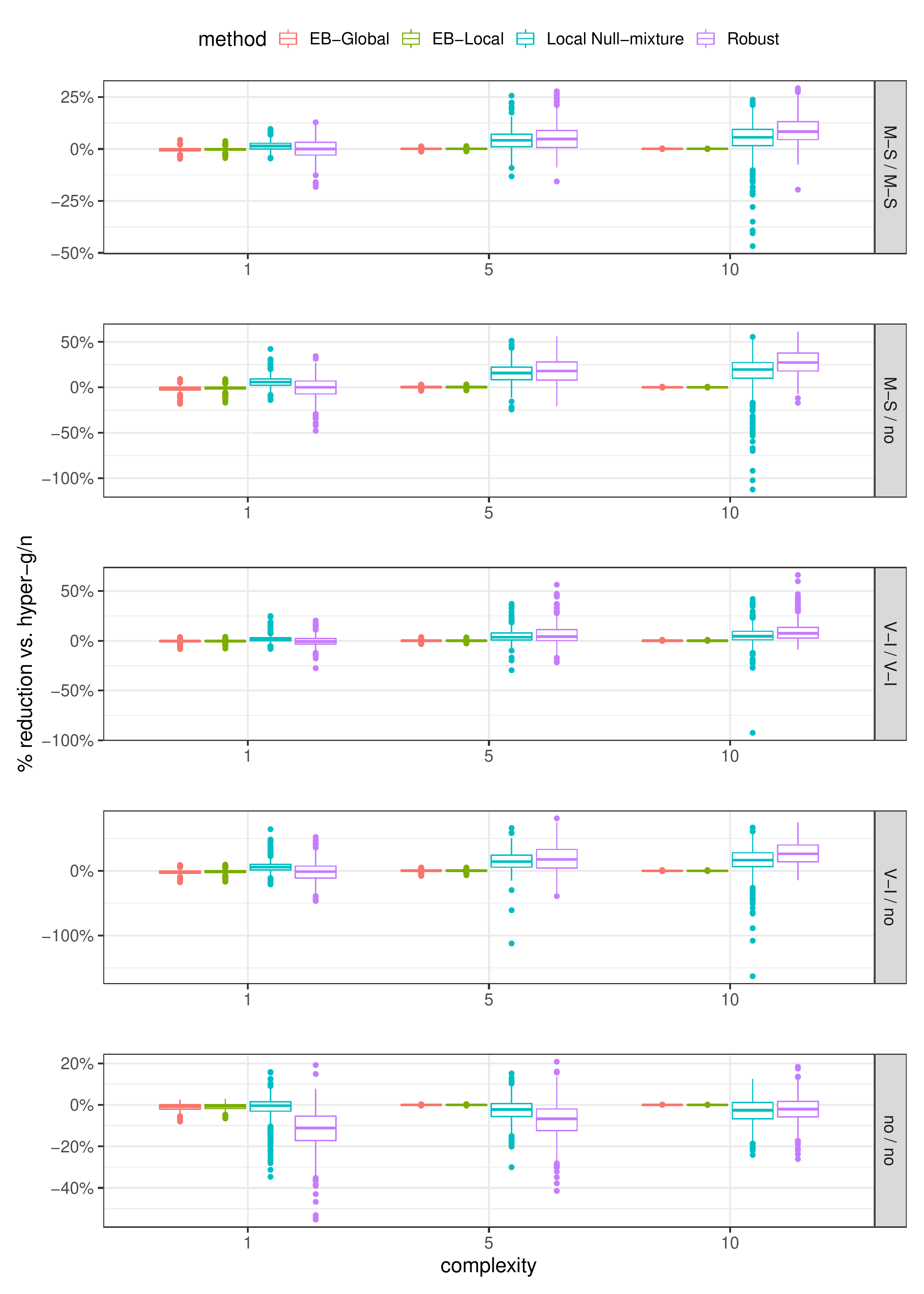}
\end{center}
\caption{Percent reduction in MSPE relative to
  the model-averaged hyper-$g/n$ method for the simulation study
  with $p=10$ predictors under
  complexity settings (the number of non-zero coefficients) 1,
  5 and 10. Each row corresponds to a contamination scheme,
  labeled ``X / Y'', where X (Y) denotes the contamination scheme for the
  training (test) data. ``M-S'' denotes mean-shift contamination, ``V-I''
  denotes variance-inflation contamination, and ``no'' denotes no
  contamination.
  }
\label{fig_MSPE_p10}
\end{figure}

We performed a second simulation study to demonstrate how the
predictive performance of the local null-mixture method scales
in comparison to the other methods as $p$ increases. The setup
is the same as above but now with double the
number of predictors ($p = 10$) and three complexity settings
(1, 5 and 10). The values of the true coefficients $\bm{\beta} = 
(\beta_1, \ldots, \beta_{10})^T$ under each of the three complexity
settings are given in the right-hand side of Table~\ref{tab:complexity_level}.
The results, summarized graphically in Figure~\ref{fig_MSPE_p10},
confirm the overall features and patterns uncovered by the first
simulation study.

\section{Crime Data}\label{sec:crime}
To evaluate the out-of-sample predictive performances of several of
the methods discussed in this article in a real data analysis setting,
we consider the crime data reported in \citet{agresti1970statistical}.
The methods we evaluate are a subset of those considered in the
simulation study (Hyper-$g/n$, EB-Local, EB-Global, and Local
Null-mixture), and we again use the Beta-Binomial$(1,1)$ 
prior over the model space.

  For each US state and the District of Columbia, the data comprise a
  dependent variable, ``violent crimes per 100,000 people,'' to be
  modeled in terms of eight socio-economic regressors.  Following a
  preliminary exploratory analysis we took a logarithmic
  transformation of the dependent variable to remedy the skewness of
  the observed counts.  In addition, the exploratory analysis
  uncovered positive correlations between the transformed dependent
  variable and some of the regressors as well as between some of the
  regressors themselves.
  Diagnostics for the full linear regression model
  revealed that the observations for Hawaii (HI) and the District of
  Columbia (DC) are the most highly influential and that a normality
  assumption for the errors in the full model might not hold.  We also
  examined residual plots produced by applying various BMA procedures
  to the data and found that the residuals for the three traditional
  methods EB-Local, EB-Global, and Hyper-$g/n$, are very similar but
  differ in several aspects from those of the local null-mixture
  method (e.g., the local null-mixture method produces a larger
  residual for Alaska (AK) and a smaller one for DC).

  We used $K$-fold cross-validation \citep{friedman2001elements} to
  evaluate the out-of-sample predictive performance of the various
  methods, executing the following steps: (a)~partition the
  observations at random into $K$ subsets having approximately equal
  size; (b)~conditional on the selected partition, leave out in turn
  one of the $K$ subsets as the testing data, and use the remaining
  $K-1$ subsets as the training data; (c)~apply the various BMA
  methods to the training data to make predictions on the testing
  data.

  Using the cross-validated predictions, the cross-validation error
  (CVE) of each method, is calculated as
\begin{equation} \label{eq:cv}
CV\!E(\hat{f},\text{method},K,\mathcal{P})=\frac{1}{n}
\sum_{i=1}^n\left(Y_i-\hat{f}^{\setminus\kappa(i)}(\bm{X_i},
  \text{method})\right)^2,   
\end{equation}
where 
$n$ is the number of observations in the data set, $K$ is the
  number of sets in the   
  selected partition $\mathcal{P}$, 
$\kappa(i)$ denotes the held-out
element of the partition $\mathcal{P}$ to which 
observation $i$ belongs, and
$\hat{f}^{\setminus\kappa(i)}(\bm{X_i},\text{method})$ denotes
  the fitted value at $\bm{X_i}$ produced by a given method when
  the $\kappa(i)$-th element of the partition is removed. 

  For given $K$, the CVE value varies from partition to partition and,
  assuming a uniform distribution over partitions, we can define an
  expected cross-validation error (ECVE).  When $K=51$, there is only
  one partition and the ECVE can be computed exactly using
  Equation~\eqref{eq:cv}.  For smaller values of $K$, when there are
  too many partitions to compute the exact expectation, we repeat the
  $K$-fold cross-validation procedure $T$ times and approximate the
  ECVE by
\begin{equation}
\label{est_ECVE}
\widehat{ECV\!E}(\hat{f},\text{method},K)=\frac{1}{T}
\sum_{t=1}^T\left(CV\!E(\hat{f},\text{method},K,\mathcal{P}^{(t)})\right), 
\end{equation}
where $\mathcal{P}^{(t)}$ represents a partition selected at random in
repetition $(t)$.  

The cross-validation results for values of $K$ equal to 51, 25, and 10
are summarized in Table~\ref{chp5_crime_data}.  The EB-Local and
EB-Global methods always perform comparably to one another and
slightly better than the Hyper-$g/n$ method.  The Local Null-mixture
method clearly outperforms all other methods for $K$ equal to 51, 
is noticeably better for $K$ equal to 25, 
and is only slightly worse for $K$ equal to 10.

\begin{table}[t!]
\centering
  \begin{tabular}{l c c c c c c}
 \hline\hline
 \multicolumn{1}{c}{}&
 \multicolumn{2}{c}{$K=51$}&
 \multicolumn{2}{c}{$K=25$}&
 \multicolumn{2}{c}{$K=10$}
 \\  \hline
    Method & $ECV\!E$ & \% Red. &  $\widehat{ECV\!E}$ & \% Red. &
        $\widehat{ECV\!E}$ & \% Red. \\
   \hline
 Hyper-$g/n$ &0.204 &  $\cdot$ & 0.204 & $\cdot$ & 0.209 & $\cdot$ \\
 EB-Local & 0.202 & 0.94 & 0.203 & 0.93 & 0.207 & 0.91\\
 EB-Global & 0.202  & 1.17 & 0.202 & 1.16 & 0.207 & {\bf 1.11} \\
 Local Null-mixture & 0.174   &  {\bf 14.73} & 0.197 & {\bf 3.87} & 0.218 &-4.40\\
  \hline\hline
  \end{tabular}
  \caption{For the crime data, $ECV\!E$ and $\widehat{ECV\!E}$ for
    various BMA methods and the realized percent reduction in their
    values compared to those obtained using the Hyper-$g/n$ method.
    The largest realized percent reduction for each cross-validation
    setting $K$ appears in boldface.}
\label{chp5_crime_data}
\end{table}

The two most influential cases
(DC and HI) have a similar impact on
the full model.
Apparently, the 
Local Null-mixture method benefits from the inclusion of one of these
two cases in the training set and of the other in the
testing set.  This happens with probability one when $K$
equals 51 and with decreasing probability as $K$ becomes larger.

\section{Discussion}\label{sec:discussion}
We developed BMA model prediction methodology for common types of
regression model misfit, in particular the presence of influential
observations and non-constant residual variance.  The methodology is
motivated by the view that, typically, good predictive performance
cannot be attained unless the model residuals are well behaved.  The
regression models under consideration make use of variants on
Zellner's $g$ prior.  By studying the impact of various forms of model
misfit on BMA predictions in simple situations we were able to
identify prescriptive guidelines for ``tuning" Zellner's $g$ prior to
obtain optimal predictions. The methodology can be thought of as an
``empirical Bayes" approach to modeling, as the data help to inform
the specification of the prior in an attempt to attenuate the negative
impact of model misfit.  The methods described in the paper can be
extended to other types of model misfit, such as non-linearity of the
mean function (or model under-fit), and can be implemented with
different adaptations of robust $g$-prior specification, as
illustrated in \citet{wang2016empirical}.

In modern applications, especially when there are many potential
predictor variables, analysts do not have the time (or, due to the
size of the problem, are not able) to interactively investigate the
quality of fitted models, hampering one's ability to manually
attenuate the impact of model misfit through either model revision or
prior tuning. In such complex situations, the guidelines developed in
the simple examples considered in this paper can motivate automatic or
semi-automatic procedures that provide some insurance against Bayesian
predictions that are unduly impacted by gross model specification.

Standard implementations of BMA analyses based on Zellner's $g$ prior
typically make the simplifying assumption that the prior mean of the
regression coefficients is zero, leading to shrinkage of the
least squares estimate toward the origin, which may be sub-optimal in
the presence of model misfit. By developing BMA strategies that take
advantage of the full generality of the prior as proposed by Zellner,
in particular by allowing for a non-zero prior mean, we are able to
achieve shrinkage toward points other than the origin, which can have
the effect of attenuating the impact of model misfit on prediction.
There is little doubt that our proposed methodology may not be as good
as the best possible human analysis (with the understanding, of
course, that the quality of such an analysis depends on the skills of
the individual analyst), but we have demonstrated through theoretical
arguments and empirical investigations that our methodology performs
better than routine implementations of BMA that do not account for
model misfit.

In our methodology, the tuning of the prior distribution is obtained
by considering theoretical properties that should be enjoyed by the
optimal fit (mainly, the parallel condition) of the various models in
the BMA ensemble. This approach should lead to well-behaved residuals
for each of the individual models. An alternative approach to
fine-tuning the prior distributions could focus on the realized
residuals of the model-averaged fit. This would be a more empirical
approach requiring an objective function that assesses quantitatively
the quality of the realized residuals coupled with a feasible
computational approach for sequentially updating the prior parameters
to improve such an objective function. An appeal of such an approach
is the potential for the development of dynamic, graphical diagnostics
that are closely related to the traditional diagnostics for linear
regression that are familiar to most users.

Finally, extension of our methods to
situations involving very large data sets may require the
development of algorithms that implement computational shortcuts for
the determination of the prior tuning parameters. A guiding direction
for such work would be to develop shortcuts that, while speeding up
computation, do not unduly compromise the Bayesian optimality
guarantees implicit in our framework.

\paragraph{Acknowledgments.}
This work was supported by the National Science Foundation under
awards No.~DMS-1310294, No.~SES-1424481, and No.~SES-1921523.  The
manuscript is based on the third author's Ph.D. dissertation completed
at The Ohio State University.

\bibliographystyle{natbib}
\bibliography{references-ebma}

\end{document}